\documentclass[twoside,leqno]{article}
\usepackage{hyperref}
\usepackage{amsmath}
\usepackage{amsthm}
\usepackage{amssymb}
\usepackage{amscd}
\usepackage{latexsym}
\usepackage{amsfonts}

\input xy
\xyoption{all} \tolerance=500

\hypersetup{
pdfauthor={Luca Vitagliano, Katarzyna Grabowska},
pdftitle={Triple},
pdfborder={0 0 0},
colorlinks=true
}

=1in \textwidth 15.6cm \topmargin 15mm\textheight23.1cm \advance\topmargin by -\headheight\advance\topmargin by
-\headsep

\numberwithin{equation}{section} \allowdisplaybreaks

\theoremstyle{definition}

\newtheorem{example}{Example} [section]

\begin{document}


\font\black=cmbx10 \font\sblack=cmbx7 \font\ssblack=cmbx5 \font\blackital=cmmib10  \skewchar\blackital='177
\font\sblackital=cmmib7 \skewchar\sblackital='177 \font\ssblackital=cmmib5 \skewchar\ssblackital='177
\font\sanss=cmss10 \font\ssanss=cmss8 
\font\sssanss=cmss8 scaled 600 \font\blackboard=msbm10 \font\sblackboard=msbm7 \font\ssblackboard=msbm5
\font\caligr=eusm10 \font\scaligr=eusm7 \font\sscaligr=eusm5 \font\blackcal=eusb10 \font\fraktur=eufm10
\font\sfraktur=eufm7 \font\ssfraktur=eufm5 \font\blackfrak=eufb10

\font\bsymb=cmsy10 scaled\magstep2
\def\all#1{\setbox0=\hbox{\lower1.5pt\hbox{\bsymb
       \char"38}}\setbox1=\hbox{$_{#1}$} \box0\lower2pt\box1\;}
\def\exi#1{\setbox0=\hbox{\lower1.5pt\hbox{\bsymb \char"39}}
       \setbox1=\hbox{$_{#1}$} \box0\lower2pt\box1\;}

\def\mi#1{{\fam1\relax#1}}
\def\tx#1{{\fam0\relax#1}}

\newfam\bifam
\textfont\bifam=\blackital \scriptfont\bifam=\sblackital \scriptscriptfont\bifam=\ssblackital
\def\bi#1{{\fam\bifam\relax#1}}

\newfam\blfam
\textfont\blfam=\black \scriptfont\blfam=\sblack \scriptscriptfont\blfam=\ssblack
\def\rbl#1{{\fam\blfam\relax#1}}

\newfam\bbfam
\textfont\bbfam=\blackboard \scriptfont\bbfam=\sblackboard \scriptscriptfont\bbfam=\ssblackboard
\def\bb#1{{\fam\bbfam\relax#1}}

\newfam\ssfam
\textfont\ssfam=\sanss \scriptfont\ssfam=\ssanss \scriptscriptfont\ssfam=\sssanss
\def\sss#1{{\fam\ssfam\relax#1}}

\newfam\clfam
\textfont\clfam=\caligr \scriptfont\clfam=\scaligr \scriptscriptfont\clfam=\sscaligr
\def\cl#1{{\fam\clfam\relax#1}}

\newfam\frfam
\textfont\frfam=\fraktur \scriptfont\frfam=\sfraktur \scriptscriptfont\frfam=\ssfraktur
\def\fr#1{{\fam\frfam\relax#1}}

\def\cb#1{\hbox{$\fam\gpfam\relax#1\textfont\gpfam=\blackcal$}}

\def\hpb#1{\setbox0=\hbox{${#1}$}
    \copy0 \kern-\wd0 \kern.2pt \box0}
\def\vpb#1{\setbox0=\hbox{${#1}$}
    \copy0 \kern-\wd0 \raise.08pt \box0}

\def\pmb#1{\setbox0\hbox{${#1}$} \copy0 \kern-\wd0 \kern.2pt \box0}
\def\pmbb#1{\setbox0\hbox{${#1}$} \copy0 \kern-\wd0
      \kern.2pt \copy0 \kern-\wd0 \kern.2pt \box0}
\def\pmbbb#1{\setbox0\hbox{${#1}$} \copy0 \kern-\wd0
      \kern.2pt \copy0 \kern-\wd0 \kern.2pt
    \copy0 \kern-\wd0 \kern.2pt \box0}
\def\pmxb#1{\setbox0\hbox{${#1}$} \copy0 \kern-\wd0
      \kern.2pt \copy0 \kern-\wd0 \kern.2pt
      \copy0 \kern-\wd0 \kern.2pt \copy0 \kern-\wd0 \kern.2pt \box0}
\def\pmxbb#1{\setbox0\hbox{${#1}$} \copy0 \kern-\wd0 \kern.2pt
      \copy0 \kern-\wd0 \kern.2pt
      \copy0 \kern-\wd0 \kern.2pt \copy0 \kern-\wd0 \kern.2pt
      \copy0 \kern-\wd0 \kern.2pt \box0}

\def\cdotss{\mathinner{\cdotp\cdotp\cdotp\cdotp\cdotp\cdotp\cdotp
        \cdotp\cdotp\cdotp\cdotp\cdotp\cdotp\cdotp\cdotp\cdotp\cdotp
        \cdotp\cdotp\cdotp\cdotp\cdotp\cdotp\cdotp\cdotp\cdotp\cdotp
        \cdotp\cdotp\cdotp\cdotp\cdotp\cdotp\cdotp\cdotp\cdotp\cdotp}}


\font\frak=eufm10 scaled\magstep1 \font\fak=eufm10 scaled\magstep2 \font\fk=eufm10 scaled\magstep3
\font\scriptfrak=eufm10 \font\tenfrak=eufm10

\mathchardef\za="710B  
\mathchardef\zb="710C  
\mathchardef\zg="710D  
\mathchardef\zd="710E  
\mathchardef\zve="710F 
\mathchardef\zz="7110  
\mathchardef\zh="7111  
\mathchardef\zvy="7112 
\mathchardef\zi="7113  
\mathchardef\zk="7114  
\mathchardef\zl="7115  
\mathchardef\zm="7116  
\mathchardef\zn="7117  
\mathchardef\zx="7118  
\mathchardef\zp="7119  
\mathchardef\zr="711A  
\mathchardef\zs="711B  
\mathchardef\zt="711C  
\mathchardef\zu="711D  
\mathchardef\zvf="711E 
\mathchardef\zq="711F  
\mathchardef\zc="7120  
\mathchardef\zw="7121  
\mathchardef\ze="7122  
\mathchardef\zy="7123  
\mathchardef\zf="7124  
\mathchardef\zvr="7125 
\mathchardef\zvs="7126 
\mathchardef\zf="7127  
\mathchardef\zG="7000  
\mathchardef\zD="7001  
\mathchardef\zY="7002  
\mathchardef\zL="7003  
\mathchardef\zX="7004  
\mathchardef\zP="7005  
\mathchardef\zS="7006  
\mathchardef\zU="7007  
\mathchardef\zF="7008  
\mathchardef\zW="700A  

\newcommand{\be}{\begin{equation}}
\newcommand{\ee}{\end{equation}}
\newcommand{\ra}{\rightarrow}
\newcommand{\lra}{\longrightarrow}
\newcommand{\bea}{\begin{eqnarray}}
\newcommand{\eea}{\end{eqnarray}}
\newcommand{\beas}{\begin{eqnarray*}}
\newcommand{\eeas}{\end{eqnarray*}}
\def\*{{\textstyle *}}
\newcommand{\R}{{\mathbb R}}
\newcommand{\T}{{\mathbb T}}
\newcommand{\C}{{\mathbb C}}
\newcommand{\unit}{{\mathbf 1}}
\newcommand{\SL}{SL(2,\C)}
\newcommand{\Sl}{sl(2,\C)}
\newcommand{\SU}{SU(2)}
\newcommand{\su}{su(2)}
\def\ssT{\sss T}
\newcommand{\G}{{\goth g}}
\newcommand{\D}{{\rm d}}
\newcommand{\Df}{{\rm d}^\zF}
\newcommand{\de}{\,{\stackrel{\rm def}{=}}\,}
\newcommand{\we}{\wedge}
\newcommand{\nn}{\nonumber}
\newcommand{\ot}{\otimes}
\newcommand{\s}{{\textstyle *}}
\newcommand{\ts}{T^\s}
\newcommand{\oX}{\stackrel{o}{X}}
\newcommand{\oD}{\stackrel{o}{D}}
\newcommand{\obD}{\stackrel{o}{\bD}}
\newcommand{\pa}{\partial}
\newcommand{\ti}{\times}
\newcommand{\A}{{\cal A}}
\newcommand{\Li}{{\cal L}}
\newcommand{\ka}{\mathbb{K}}
\newcommand{\find}{\mid}
\newcommand{\ad}{{\rm ad}}
\newcommand{\rS}{]^{SN}}
\newcommand{\rb}{\}_P}
\newcommand{\p}{{\sf P}}
\newcommand{\h}{{\sf H}}
\newcommand{\X}{{\cal X}}
\newcommand{\I}{\,{\rm i}\,}
\newcommand{\rB}{]_P}
\newcommand{\Ll}{{\pounds}}
\def\lna{\lbrack\! \lbrack}
\def\rna{\rbrack\! \rbrack}
\def\rnaf{\rbrack\! \rbrack_\zF}
\def\rnah{\rbrack\! \rbrack\,\hat{}}
\def\lbo{{\lbrack\!\!\lbrack}}
\def\rbo{{\rbrack\!\!\rbrack}}
\def\lan{\langle}
\def\ran{\rangle}
\def\zT{{\cal T}}
\def\tU{\tilde U}
\def\ati{{\stackrel{a}{\times}}}
\def\sti{{\stackrel{sv}{\times}}}
\def\aot{{\stackrel{a}{\ot}}}
\def\sati{{\stackrel{sa}{\times}}}
\def\saop{{\stackrel{sa}{\op}}}
\def\bwa{{\stackrel{a}{\bigwedge}}}
\def\sv op{{\stackrel{sv}{\oplus}}}
\def\saot{{\stackrel{sa}{\otimes}}}
\def\cti{{\stackrel{cv}{\times}}}
\def\cop{{\stackrel{cv}{\oplus}}}
\def\dra{{\stackrel{\xd}{\ra}}}
\def\bdra{{\stackrel{\bd}{\ra}}}
\def\bAff{\mathbf{Aff}}
\def\Aff{\sss{Aff}}
\def\bHom{\mathbf{Hom}}
\def\Hom{\sss{Hom}}
\def\bt{{\boxtimes}}
\def\sot{{\stackrel{sa}{\ot}}}
\def\bp{{\boxplus}}
\def\op{\oplus}
\def\bwak{{\stackrel{a}{\bigwedge}\!{}^k}}
\def\aop{{\stackrel{a}{\oplus}}}
\def\ix{\operatorname{i}}
\def\V{{\cal V}}
\def\cD{{\cal D}}
\def\cC{{\cal C}}
\def\cE{{\cal E}}
\def\cL{{\cal L}}
\def\cN{{\cal N}}
\def\cR{{\cal R}}
\def\cJ{{\cal J}}
\def\cT{{\cal T}}
\def\cH{{\cal H}}
\def\bA{\mathbf{A}}
\def\bI{\mathbf{I}}
\def\wh{\widehat}
\def\wt{\widetilde}
\def\ol{\overline}
\def\ul{\underline}
\def\Sec{\sss{Sec}}
\def\Lin{\sss{Lin}}
\def\ader{\sss{ADer}}
\def\ado{\sss{ADO^1}}
\def\adoo{\sss{ADO^0}}
\def\AS{\sss{AS}}
\def\bAS{\sss{AS}}
\def\bLS{\sss{LS}}
\def\bAP{\sss{AV}}
\def\bLP{\sss{LP}}
\def\AP{\sss{AP}}
\def\LP{\sss{LP}}
\def\LS{\sss{LS}}
\def\Z{\mathbf{Z}}
\def\oZ{\overline{\bZ}}
\def\oA{\overline{\bA}}
\def\cim{{C^\infty(M)}}
\def\de{{\cal D}^1}
\def\la{\langle}
\def\ran{\rangle}
\def\by{{\bi y}}
\def\bs{{\bi s}}
\def\bc{{\bi c}}
\def\bd{{\bi d}}
\def\bh{{\bi h}}
\def\bD{{\bi D}}
\def\bY{{\bi Y}}
\def\bX{{\bi X}}
\def\bL{{\bi L}}
\def\bV{{\bi V}}
\def\bW{{\bi W}}
\def\bS{{\bi S}}
\def\bT{{\bi T}}
\def\bC{{\bi C}}
\def\bE{{\bi E}}
\def\bF{{\bi F}}
\def\bP{{\bi P}}
\def\bp{{\bi p}}
\def\bz{{\bi z}}
\def\bZ{{\bi Z}}
\def\bq{{\bi q}}
\def\bQ{{\bi Q}}
\def\bx{{\bi x}}

\def\sA{{\sss A}}
\def\sC{{\sss C}}
\def\sD{{\sss D}}
\def\sG{{\sss G}}
\def\sH{{\sss H}}
\def\sI{{\sss I}}
\def\sJ{{\sss J}}
\def\sK{{\sss K}}
\def\sL{{\sss L}}
\def\sO{{\sss O}}
\def\sP{{\sss P}}
\def\sPh{{\sss P\sss h}}
\def\sT{{\sss T}}
\def\sv {{\sss V}}
\def\sR{{\sss R}}
\def\sS{{\sss S}}
\def\sE{{\sss E}}
\def\sF{{\sss F}}
\def\st{{\sss t}}
\def\sg{{\sss g}}
\def\sx{{\sss x}}
\def\sv {{\sss v}}
\def\sw{{\sss w}}
\def\sQ{{\sss Q}}
\def\sj{{\sss j}}
\def\sq{{\sss q}}
\def\sd{{\sss d}}
\def\xa{\tx{a}}
\def\xc{\tx{c}}
\def\xd{\tx{d}}
\def\xD{\tx{D}}
\def\xV{\tx{V}}
\def\xF{\tx{F}}
\def\dt{\xd_{\sss T}}
\def\vt{\textsf{v}_{\sss T}}
\def\vta{\operatorname{v}_\zt}
\def\vtb{\operatorname{v}_\zp}
\def\cM{\cal M}
\def\cN{\cal N}
\def\cD{\cal D}
\def\ug{\ul{\zg}}
\def\sTn{\stackrel{\scriptscriptstyle n}{\textstyle\sT}\!}
\def\sTd{\stackrel{\scriptscriptstyle 2}{\textstyle\sT}\!}
\def\stn{\stackrel{\scriptscriptstyle n}{\textstyle\st}\!}
\def\std{\stackrel{\scriptscriptstyle 2}{\textstyle\st}\!}
\newdir{ (}{{}*!/-5pt/@^{(}}



\setcounter{page}{1} \thispagestyle{empty}


\bigskip

\bigskip


\title{Tulczyjew Triples in Higher Derivative Field Theory
}

\author{
Katarzyna  Grabowska\thanks{Research founded by the  Polish National Science Centre grant under the contract number DEC-2012/06/A/ST1/00256.}\\
{\it Division of Mathematical Methods in Physics, University of Warsaw} \\
{\it Ho\.za 69, 00-681 Warszawa, Poland} \\ \\
Luca Vitagliano  \\
{\it DipMat, Universit\`a degli Studi di Salerno, and}\\
{\it Istituto Nazionale di Fisica Nucleare, GC Salerno} \\
{\it Via Giovanni Paolo II, n${}^{\circ}$ 123, 84084 Fisciano (SA) Italy}
}

\date{}
\maketitle
\begin{abstract}
The geometrical structure known as Tulczyjew triple has been used with success in analytical mechanics and first order field theory to describe a wide range of physical systems, including Lagrangian/Hamiltonian systems with constraints and/or sources, or with singular Lagrangian. Starting from the first principles of the variational calculus, we derive Tulczyjew triples for classical field theories of arbitrary high order, i.e.~depending on arbitrarily high derivatives of the fields. A first triple appears as the result of considering higher order theories as first order ones with configurations being constrained to be \emph{holonomic} jets. A second triple is obtained after a reduction procedure aimed at getting rid of nonphysical degrees of freedom. This picture we present is fully covariant and complete: it contains both Lagrangian and Hamiltonian formalisms, in particular the Euler-Lagrange equations. Notice that the required Geometry of jet bundles is affine (as opposed to the linear Geometry of the tangent bundle). Accordingly, the notions of affine duality and affine phase space play a distinguished role in our picture. In particular the Tulczyjew triples in this paper consist of morphisms of double affine-vector bundles which, moreover, preserve suitable presymplectic structures.
\end{abstract}

\section{Introduction}

\subsection{Variational calculus in Statics} \label{sec:1.1}
From a mathematical point of view, calculus of variations is a theory providing tools for finding extremals, or stationary points, of functionals, i.e.~maps from a set of functions to real numbers. Using calculus of variations, one may find, for example, differential equations for curves of the shortest length connecting two points, or surfaces of minimal area spanning a given frame. In physics the same mathematical tools can be used to formulate principles of least action leading to Euler-Lagrange equations both in mechanics and field theory. In our paper we shall adopt a different point of view based on ideas of W.~M.~Tulczyjew, as presented in his numerous works and lectures (see, for instance the book \cite{Tu3} and papers \cite{Tu6,Tu7,Tu9,Tu10}).

In the Tulczyjew approach a physical system is studied through its response to interactions. This can be explained in a natural way in Statics since all mathematical objects that are used there have direct physical interpretations. Take a static system $S$, and suppose that the set of configurations of $S$ is a smooth manifold $Q$. The system $S$ can be probed by changing its configuration in a {\it quasi-static} way,
i.e.~slowly enough to produce negligible dynamical effects. The changing of configurations is called {\it a process} and is represented mathematically
by a one-dimensional smooth oriented submanifold of $Q$ with boundary. It may happen that not all the processes are admissible. In such a case we say that the system is constrained. We assume that we can estimate the {\it cost} of every process.  All the information about the system is therefore encoded in the following three objects: the configuration manifold $Q$, the set of all admissible processes, and the cost function that assigns a real number to every process. The cost function should fulfill some additional conditions, e.g.~it should be additive in the sense that if we break a process into two subprocesses, then the cost of the whole process should be equal to the sum of the costs of the two subprocesses. Usually one also assumes that the cost function is local, i.e.~for each process, it is an integral of a certain positively homogeneous function $W$ on ${\mathsf T} Q$ or, in case of a constrained system, on some subset $\Delta\subset{\mathsf T} Q$. Vectors tangent to admissible processes are called {\it admissible virtual displacements}. The set $\Delta$ should be positively homogeneous since admissible processes are a priori unparameterized.

In Statics, one is usually interested in equilibrium configurations of isolated systems, as well as systems interacting (with other static systems). A point $q\in Q$ is an {\it equilibrium point} if, for all ``short enough'' processes starting in $q$, the cost function is non-negative. The first-order necessary condition for an equilibrium point $q$ is $W(\delta q)\geq 0$ for all vectors $\delta q\in \Delta\cap{\mathsf T}_qQ$. Interactions between systems are described by composite systems. One can compose two systems that have the same configuration space $Q$. The composite system is then described by the intersection of admissible processes and the sum $W=W_1+W_2$ of the cost functions $W_1$ and $W_2$ of the two systems. We would like to know how does a system $S_1$ interact with any possible other system ($S_2$) which means that we would like to know all equilibrium points of all possible composite systems (containing $S_1$). However, making a list of all composite systems and their equilibrium points is not an efficient way of describing $S_1$. A more efficient way is discussed below.

There is a distinguished class of systems called {\it regular}, for which all the processes are admissible and the function $W$ is the differential of a given function $U:Q\rightarrow {\mathbb R}$, called the {\it potential}. Thus, restrict to the composition of $S_1$ with a regular system. The condition for an equilibrium point of the composite system reads now $W_1(\delta q)-\langle{\mathrm d} U,\delta q\rangle\geq 0$, i.e.~$W_1(\delta q)\geq\langle{\mathrm d} U,\delta q\rangle$ (the presence of a minus sign is just a matter of conventions). Let us note that a regular system at a point $q$ is represented by a covector $\varphi={\mathrm d} U(q)$ called a {\it force}. We may now make a list $\mathcal{C}_q$ of all forces in equilibrium with our system at a point $q$. The subset $\mathcal{C}=\bigcup_{q\in Q} \mathcal{C}_q$ of ${\mathsf T}^\ast Q$ is called a {\it constitutive set}. Now, given two systems with the same configuration manifold and constitutive sets $\mathcal{C}^1$ and  $\mathcal{C}^2$ we can answer the question whether or not $q$ is an equilibrium point for the composite system. This happens precisely when $\mathcal{C}^1_q\cap\mathcal{C}^2_q\neq\varnothing$. The correspondence between cost functions $W$ and the constitutive sets $\mathcal{C}$ is known as the {\it Fenchel transformation} and considered within convex analysis. If $W$ is convex then $\mathcal{C}$ contains a full information about $W$, i.e.~$W$ can be recovered from $\mathcal{C}$. In any case, all the information we need about our system is encoded by the constitutive set. For a regular system with potential $U$ the constitutive set is $\mathcal{C}={\mathrm d} U(Q)$. It is then a Lagrangian submanifold (generated by $U$) in ${\mathsf T}^\ast Q$. Notice that a regular system is in equilibrium without external forces iff its configuration is an extremal for the potential $U$.

The above ideas apply efficiently to other theories as mechanics or field theory as well. To see this we shall specify a configuration space $Q$, a set of processes (or at least infinitesimal processes), the set of functions on $Q$ (to define regular systems), the set of covectors ${\mathsf T}^\ast Q$ (to define constitutive sets). It is not always obvious how to do this since in many interesting situations $Q$ is not a manifold any more. The main aim of the present paper is showing how things work in the case of \emph{higher derivative field theory}.

\subsection{Variational calculus in mechanics}\label{sub:12}

In this section we shall concentrate on the main ideas leading to the classical Tulczyjew triple in mechanics. We briefly present two ``regimes'' of the theory: mechanics on a finite time interval and mechanics on an infinitesimal one. Analyzing the first regime allows to identify appropriate mathematical descriptions of physical quantities, while analyzing the second one provides phase equations and the Tulczyjew triple itself. We refer to \cite{Tu3} for details.

Let $M$ be the manifold of positions of a mechanical system. Thus, for motions concentrated in a finite time interval $[t_0,t_1]$ a {\it configuration} $q$ is a smooth path $q:[t_0, t_1]\rightarrow M.$ The set of all configurations will be denoted by $Q$. Since $Q$ is not a finite-dimensional manifold, it is not obvious a priori what are processes in $Q$, smooth functions on $Q$, and tangent vectors to $Q$. We adopt the following (natural) definitions. They are well enough for our purposes. Smooth functions on $Q$ are action functionals associated to Lagrangians $L:{\mathsf T} M\to{\mathbb R}$ by the usual formula $S(q)=\int_{t_0}^{t_1} L(\dot q)dt.$ Smooth, parameterized processes, or curves in $Q$, are smooth maps $\chi:{\mathbb R}^2\supset I\times [t_0, t_1]\rightarrow M$ where $I$ is some neighborhood of zero in ${\mathbb R}$. This means that at a value $s \in I$ of the parameter, the process reaches a configuration given by the path $t\mapsto \chi(s,t)$. Notice that, with these definitions, the composition of a process with a function is a smooth function ${\mathbb R}\ni s\mapsto S(\chi(s,\cdot))\in{\mathbb R}$. Having smooth curves and smooth functions on $Q$ one can define tangent and cotangent vectors as suitable equivalence classes.

A {\it tangent vector to $Q$} is an equivalence class of processes with respect to the obvious equivalence relation. Namely, two processes $\chi_1$ and $\chi_2$ are equivalent if they have the same value at $s=0$ and, for all smooth functions $S$, $(S\circ\chi_1)'(0)=(S\circ\chi_2)'(0)$, where a prime ``$(\cdot)^\prime$'' denotes derivative with respect to $s$. The equivalence class of a process $\chi$ will be temporarily denoted by $[\chi]$.  We say that $[\chi]$ is tangent at $q \in Q$ iff $\chi(0,t) = q(t)$. A {\it tangent covector to $Q$} is an equivalence class of pairs $(q,S)$ (with $q \in Q$ and $S$ a smooth function) with respect to the obvious equivalence relation. Namely, two pairs $(q_1,S_1)$ and $(q_2,S_2)$ are equivalent if $q_1=q_2$ and, for all smooth curves $\chi$ such that $\chi(0,\cdot)=q_1=q_2$, we have $(S\circ\chi_1)'(0)=(S\circ\chi_2)'(0)$. The equivalence class of a pair $(q,S)$ will be denoted by ${\mathrm d} S(q)$. We can pair a covector and a vector provided they are attached at the same configuration. The pairing is
\begin{equation}\label{mech:0}
\langle {\mathrm d} S(q), [\chi]\rangle =S\circ\chi(0)=\int_{t_0}^{t_1}\frac{d}{ds}_{|s=0}L(\dot\chi(s))d t.
\end{equation}
Vectors and covectors defined as equivalence classes are very abstract objects. It is very useful to describe them in alternative, and easy to use, ways, what Tulczyjew calls ``convenient representations''. The choice of convenient representations for vectors and covectors is based on integration by parts. Namely, integrating by parts in (\ref{mech:0}) we get
\begin{equation}\label{mech:1}
\langle {\mathrm d} S(q), [\gamma]\rangle=
\int_{t_0}^{t_1}\langle\mathcal{E}L(\ddot \gamma(0)), \delta q\rangle dt +\left.\frac{}{}\langle\,\mathcal{P} L(\dot \gamma(0)),\delta q\,\rangle\right|^{t_1}_{t_0}\,,
\end{equation}
where $\mathcal{E}L:{\mathsf T}^2M\to{\mathsf T}^*M$ is the Euler-Lagrange map and $\mathcal{P} L:{\mathsf T} M\to{\mathsf T}^*M$ is the vertical differential of the Lagrangian $L$. It is easy to see that the tangent vector $[\chi]$ is equivalent to (i.e.~it contains the same information as) the curve $\delta q: [t_0, t_1]\rightarrow {\mathsf T} M$, where $\delta q(t)$ is the tangent vector to the curve $s\mapsto \chi(s, t)$ at $s=0$. Similarly, the covector ${\mathrm d} S(q)$ is equivalent to the triple
$(f,p_0,p_1)$, where $f:[t_0,t_1]\rightarrow{\mathsf T}^\ast M$, $f(t)=\mathcal{E}L(\ddot q(t))$ and  $p_a\in{\mathsf T}^\ast_{q(t_a)}M$, $p_a=\mathcal{P} L(\dot q(t_a))$, $a = 0,1$. Paths in ${\mathsf T} M$ and triples $(f,p_0,p_1)$ as above are Tulczyjew {\it convenient representatives} of vectors and covectors, respectively. Correspondences $[\chi] \mapsto \delta q$ and ${\mathrm d} S (q) \mapsto (f,p_0,p_1)$ between vectors and covectors and their convenient representations are usually denoted by $\kappa$ and $\alpha$, respectively.

\emph{A mechanical system with Lagrangian $L$ is, from a static point of view, a regular system with cost function given by ${\mathrm d} S$.} Accordingly, the constitutive set is $\mathcal{C}={\mathrm d} S(Q)$. Using convenient representations one sees that $\mathcal{C}$ is actually the \emph{dynamics} of the system. More precisely, the {\it phase dynamics} of a mechanical system moving in a finite interval is the subset $\mathcal{D}$ of $\{
\text{triples }(f,p_0,p_1)\}$ defined by
$$\mathcal{D}=\alpha^{-1} (\mathcal{C})=\alpha^{-1}({\mathrm d} S(Q)),$$
i.e.,
$$\mathcal{D}=\left\{(f,p_0,p_1):\  f(t)=\mathcal{E}L(\ddot q(t)),\quad p_a=\mathcal{P} L(\dot q(t_a))\,,\ a=0,1\right\}\,.$$
Explicitly, in coordinates, $q=(x^i(t))$, $\dot q=(x^i(t),\dot x^j(t))$, and we have
$$f_i(t)=\frac{\partial L}{\partial x^i}(\dot q(t))-\frac{{\mathrm d}}{{\mathrm d} t}\left(\frac{\partial L}{\partial \dot{x}^i}(\dot q(t))\right)\,,\quad (p_a)_i=\frac{\partial L}{\partial \dot{x}^i}(\dot q(t_a))\,,\ a=0,1\,.$$
The target space ${\mathsf T}^\ast M$ of $f$ in naturally interpreted as the \emph{phase space}, i.e.~the space of \emph{momenta}.

We now pass to a different ``theoretical regime'': the one when the time interval $[t_0,t_1]$ is infinitesimally small. The appropriate notions are thus obtained in the limit $t_1-t_0 = O (dt)$. For instance, configurations are infinitesimal paths, i.e.~tangent vectors to positions. We conclude that, in the ``infinitesimal regime'', $Q = {\mathsf T} M$. In particular, the configuration space is a finite-dimensional manifold, and it is then clear what processes, i.e.~curves in $Q$, functions on $Q$, tangent and cotangent vectors are. A Lagrangian $L$ is now interpreted as a potential for the cost function ${\mathrm d} L$ of a regular system, and the constitutive set is just $\mathcal{C}={\mathrm d} L({\mathsf T} M)\subset{\mathsf T}^\ast{\mathsf T} M$. In this case, it is interesting what one gets as convenient representations of vectors and covectors. The infinitesimal version of $\kappa$ is the well-known canonical involution
\begin{equation}\label{mech:2}
\kappa_M : {\mathsf T} {\mathsf T} M \longrightarrow {\mathsf T} {\mathsf T} M.
\end{equation}

The infinitesimal version of Formula (\ref{mech:1}) reads
\begin{equation*}
\langle{\mathrm d} L, \delta\dot\chi(0,0)\rangle=\langle \mathcal{E}L(\ddot\chi(0,0)), \delta \chi(0,0)\rangle+
\frac{{\mathrm d}}{{\mathrm d} t}_{|t=0}\langle\mathcal{P}L(\dot\chi(t,0)),\delta \chi(t,0)\rangle\,.
\end{equation*}
Similarly, the infinitesimal version of $\alpha$ is the \emph{Tulczyjew isomorphism}
\begin{equation*}
\alpha_M:{\mathsf T}{\mathsf T}^\ast M \longrightarrow {\mathsf T}^\ast{\mathsf T} M\,.
\end{equation*}
Finally, the constitutive set is ${\mathrm d} L ({\mathsf T} M)$ or, when ``conveniently represented'' via $\alpha_M$:
\[
 \mathcal{D}=\alpha_M^{-1}({\mathrm d} L({\mathsf T} M)) \subset {\mathsf T}{\mathsf T}^\ast M.
\]
Since $\mathcal{D}$ is a subset of the tangent space, it can be regarded as an (implicit) first-order differential equation for curves in the phase space. Actually, it is precisely the dynamics of the system.

Now we are ready to present the Lagrangian part of the Tulczyjew triple that contains all the structure needed in the Lagrangian formulation of mechanics in the infinitesimal regime. The map $\alpha_M$ is an isomorphism of double vector bundles \cite{KU}. Both ${\mathsf T} {\mathsf T}^\ast M$ and ${\mathsf T}^\ast {\mathsf T} M$ are symplectic manifolds. The map $\alpha_M$ is also a symplectomorphism. It follows that $\mathcal{D}$ is a Lagrangian submanifold.

\begin{equation*}
\begin{array}[c]{c}
 \xymatrix@C-20pt{
{ \mathcal{D}\ }\ar@{^{(}->}[r] & {\mathsf T}{\mathsf T}^\ast M \ar[rrr]^{\alpha_{M}} \ar[dl]\ar[ddr]|!{[drr];[dl]}\hole
& & & {\mathsf T}^\ast{\mathsf T} M \ar[dl]\ar[ddr]& \\
{\mathsf T}^\ast M \ar@{=}[rrr]\ar[ddr] & & & {\mathsf T}^\ast M \ar[ddr]& & \\
 & & {\mathsf T} M  \ar@{=}[rrr]|!{[ur];[drr]}\hole \ar[dl] & & & {\mathsf T} M \ar@{=}[lll]|!{[ull];[dl]}\hole \ar[dl] \ar[ull]_-{\mathcal{P} L} \ar@/_1pc/[uul]_{{\mathrm d} L} \\
 & M \ar@{=}[rrr] & & & M &
}
\end{array}.
\end{equation*}

In the infinitesimal regime one sees that, in some cases, the dynamics $\mathcal{D}$ is the image of a vector field on ${\mathsf T}^\ast M$. Since $\mathcal{D}$ is a Lagragian submanifold this vector field should be at least locally Hamiltonian. The correspondence between functions on ${\mathsf T}^\ast M$ and Hamiltonian vector fields can be described using the following diagram:

\begin{equation*}\label{mech:7}
\begin{array}[c]{c}
 \xymatrix@C-20pt{
 & {\mathsf T}^\ast{\mathsf T}^\ast M \ar[dl]\ar[ddr]|!{[drr];[dl]}\hole
& & & {\mathsf T}{\mathsf T}^\ast M \ar[lll]_-{\beta_M} \ar[dl]\ar[ddr]& {\ \mathcal{D}}\ar@{_{(}->}[l]  \\
{\mathsf T}^\ast M \ar@{=}[rrr]\ar[ddr] & & & {\mathsf T}^\ast M \ar[ddr]& & \\
 & & {\mathsf T} M  \ar@{=}[rrr]|!{[ur];[drr]}\hole \ar[dl] & & & {\mathsf T} M\ar[dl] \ar@{=}[lll]|!{[ull];[dl]}\hole \\
 & M \ar@{=}[rrr] & & & M &
}
\end{array}.
\end{equation*}

The map $\beta_M$ is the isomorphism (of double vector bundles) determined by the canonical symplectic form $\omega_M$ on ${\mathsf T}^\ast M$, i.e.~for $X\in{\mathsf T}{\mathsf T}^\ast M$, we have $\beta_M(X)=\omega_M(\cdot, X)$. The map $\beta_M$ is an antisymplectomorphism with respect to the symplectic forms $\omega_{{\mathsf T}^\ast M}$ on ${\mathsf T}^\ast{\mathsf T}^\ast M$ and ${\mathrm d}_{{\mathsf T}}\omega_M$ on ${\mathsf T}{\mathsf T}^\ast M$  (here, $\mathsf{d}_{{\mathsf T}} \omega_M$ is the total lift of $\omega_M$). If $H$ is a function on ${\mathsf T}^\ast M$, then the
Hamiltonian vector field $X_H$ associated to $H$ is given by ${\mathrm d} H=\omega_M(\cdot, X_H)$, and we have of course, ${\mathrm d} H({\mathsf T}^\ast Q)=\beta_M(X_H({\mathsf T}^\ast Q))$. Diagram (\ref{mech:7}) is the Hamiltonian side of the Tulczyjew triple and contains all the structure needed in the Hamiltonian formulation of mechanics in the infinitesimal regime. Notice that it does not have any counterpart in the finite time interval regime.

A Lagrangian $L$ or an Hamiltonian $H$ are examples of \emph{generating objects}, i.e.~they can be used to generate a Lagrangian submanifold $\mathcal{D} \subset {\mathsf T} {\mathsf T}^\ast M$, i.e.~a dynamics. However Lagrangian submanifolds can be also generated starting from more general \emph{generating objects} using \emph{symplectic relations techniques} (see \cite{Be} for a general discussion on Lagrangian submanifolds, generating objects and symplectic relations).
The passage from Lagrangian to Hamiltonian generating objects of the dynamics is called the \emph{Legendre transformation} (not to be confused with the \emph{Legendre map} $\mathcal{P}L$). It is well known that the dynamics obtained from an hyperregular Lagrangian $L$ is the image of an Hamiltonian vector field $X_H$. In such a case  $\mathcal{D}$ has two, particularly simple, ``generating objects'', namely the functions $L:{\mathsf T} M\rightarrow {\mathbb R}$ and $H:{\mathsf T}^\ast M\rightarrow {\mathbb R}$, where $H(p)=\langle p, (\mathcal{P}L)^{-1}(p)\rangle-L((\mathcal{P}L)^{-1}(p))$. In particular we have $\mathcal{D}=\alpha_M^{-1}({\mathrm d} L({\mathsf T} M))=\beta_M^{-1}({\mathrm d} H({\mathsf T}^\ast M))$. Notice, however, that even if $L$ is not hyperregular, $\mathcal{D}= \alpha_M^{-1}({\mathrm d} L({\mathsf T} M))$ can still be generated by a suitable Hamiltonian generating object (via a suitable procedure) but not an object as simple as a function on ${\mathsf T}^\ast M$, rather a family of functions. Specifically, the ``Lagrangian bundle'' ${\mathsf T}^\ast {\mathsf T} M$ and the ``Hamiltonian bundle'' ${\mathsf T}^\ast{\mathsf T}^\ast M$ are canonically isomorphic as double vector bundles. The graph of the isomorphism $\mathcal{R}_{{\mathsf T} M} : {\mathsf T}^\ast {\mathsf T} M \rightarrow {\mathsf T}^\ast {\mathsf T}^\ast M$ is the Lagrangian submanifold generated in ${\mathsf T}^\ast ({\mathsf T} M\times{\mathsf T}^\ast M)\simeq {\mathsf T}^\ast{\mathsf T} M\times{\mathsf T}^\ast{\mathsf T}^\ast M$ by the canonical evaluation of vectors and covectors on $M$. The isomorphism $\mathcal{R}_{{\mathsf T} M}$ is an antisymplectomorphism and $\beta_M = \mathcal{R}_{{\mathsf T} M} \circ \alpha_M$. Following the rules of composing symplectic relations we get that the Lagrangian submanifold $\mathcal{R}_{TM}({\mathrm d} L({\mathsf T} M))$ is generated by a family of functions (also called a \emph{generating family})
on ${\mathsf T}^\ast M$ parameterized by elements of ${\mathsf T} M$,
\[
{\mathsf T}^\ast M\times_M {\mathsf T} M \longrightarrow {\mathbb R}, \quad (p,v) \longmapsto \tilde{H}(v,p):=L(v)-\langle p,\,v\rangle.
\]

The full Tulczyjew triple in mechanics is the diagram
\begin{equation}\label{mech:8}
\begin{array}[c]{c}
 \xymatrix@C-20pt{&&&& {\underset{\ }{\mathcal{D}}}\ar@{^{(}->}[d]&&&&\\
& {\mathsf T}^\ast{\mathsf T}^\ast M\ar[dl]\ar[ddr]|!{[drr];[dl]}\hole & & & {\mathsf T}{\mathsf T}^\ast M \ar[lll]_{\beta_{ M }}\ar[rrr]^{\alpha_{ M }} \ar[dl]\ar[ddr]|!{[drr];[dl]}\hole
& & & {\mathsf T}^\ast{\mathsf T} M \ar[dl]\ar[ddr]& \\
{\mathsf T}^\ast M \ar@/^1pc/[ur]^{{\mathrm d} H} \ar[ddr] & & & {\mathsf T}^\ast M \ar@{=}[lll]\ar@{=}[rrr]\ar[ddr] & & & {\mathsf T}^\ast M  \ar[ddr]& & \\
& & {\mathsf T} M \ar[dl] \ar@{=}[rrr]|!{[ur];[drr]}\hole & & & {\mathsf T} M \ar@{=}[lll]|!{[ull];[dl]}\hole  \ar@{=}[rrr]|!{[ur];[drr]}\hole \ar[dl] & & & {\mathsf T} M \ar@{=}[lll]|!{[ull];[dl]}\hole \ar[dl] \ar@/_1pc/[uul]_{{\mathrm d} L} \\
&  M  & & & M \ar@{=}[lll]\ar@{=}[rrr] & & &  M  &
}
\end{array}.
\end{equation}

Using the structure encoded in the Tulczyjew triple, one can describe more complicated mechanical systems than those usually treated in the traditional Lagrangian and Hamiltonian mechanics. In geometrical optics, for example, one finds systems for which one needs more general generating object on the Lagrangian side while in relativistic mechanics, one needs generating families on the Hamiltonian side, (see for instance \cite{TU1}). Finally, Diagram (\ref{mech:8}) shows also that, from the mathematical point of view, Hamiltonian and Lagrangian mechanics are equivalent only if we agree to use the most general generating objects. However, one should keep in mind that Lagrangian mechanics has variational origin and comes from the finite time interval regime after passing to the suitable limit. On the other hand, Hamiltonian mechanics comes from the theory of generating objects of Lagrangian submanifolds and symplectic relations and does not have a finite time interval counterpart.

In classical field theory, one is often interested in Euler-Lagrange PDEs, i.e.~those PDEs coming from a variational principle. Let us recall the geometric definition of a variational principle. Let $D \subset M$ be a bounded domain in the $m$-dimensional space-time $M$. Informally, a variational principle on fields described as sections of a bundle $E$ is given by an action functional
\begin{equation}
\{ \text{sections $\sigma$ of $E$} \} \mapsto \int_D L(x,\sigma(x),\nabla \sigma(x), \ldots, \nabla^{k+1} \sigma (x)), \label{Luca3}
\end{equation}
where $L$ is a \emph{Lagrangian density}, i.e.~a differential $m$-form, to be integrated on $D$, depending on a space-time point $x$, and derivatives, $\sigma(x),\nabla \sigma(x), \ldots, \nabla^{k+1} \sigma (x)$, of a section $\sigma$ of $E$ at the point $x$, up to some finite order $k+1$. From a precise, geometric point of view, a $(k+1)$-st order Lagrangian density should be understood as an $m$-form $L$ on $M$ with values in functions on the space ${\mathsf J}^{k+1}$ of $(k+1)$-st jets of sections of $E$, i.e.~a section of the line bundle ${\mathsf J}^{k+1} \times_M \Omega^m \rightarrow {\mathsf J}^{k+1}$ \cite{B,S}. An easy integration by parts shows that extremals of the action functional (\ref{Luca3}) (with respect to variation fixing the values of $\sigma$ at the boundary $\partial D$) are solutions of the Euler-Lagrange equations. The Euler-Lagrange equations are $(2k+2)$-nd order PDEs. From a geometric point of view, they consist in a submanifold $\mathcal{E} \subset {\mathsf J}^{2k+2}$ canonically associated to a given Lagrangian density $L$ \cite{S}. A $(k+1)$-st order Lagrangian density defines a $(k+1)$-st order (classical) Lagrangian field theory. The relationship between variational principles and the associated Euler-Lagrange equations, or, more generally, the calculus of variations, is rather well understood in intrinsic, geometric (and homological) terms. The main works on the topic are due to Tulczyjew \cite{Tu11} and Vinogradov \cite{V, Vbis}.  Notice, however, that the Euler-Lagrange equations do not exhaust all the relevant geometric content of a field theory even when the theory is defined by a variational principle. Fixing variations equal zero at the boundary means neglecting boundary terms which should be included in the theory. We have seen above that boundary terms in mechanics are momenta. In Electrodynamics boundary terms are related to magnetic strength and electric induction. Moreover Euler-Lagrange equations with right-hand-side equal to zero are equations for fields without sources (or external forces in mechanics). Using the Tulczyjew approach one can include both boundary terms and sources in field theory.

The main aim of this paper is twofold: (1) showing that Tulczyjew paradigms apply as well to higher order field theories, i.e.~systems whose configurations are sections of a generic bundle on a ``space-time manifold'', and whose cost functions are (in the regular case) action functionals whose Lagrangian density depends on space-time derivatives of the configurations up to arbitrarily high order; (2) showing that all the mathematical structure needed in a suitable infinitesimal regime (dynamics on an infinitesimal space-time region) is encoded by a certain (field theoretic, higher order) version of the Tulczyjew triple (\ref{mech:8}).

In \cite{G2,G} the first author made the first steps in this direction, discussing first derivative field theory, and this paper heavily relies on that one. However, higher order field theories exhibit novel features as we explain below.

\section{Mathematical background}\label{sec:2}

A convenient differential geometric setting for intrinsic aspects of partial differential equations (PDEs) in provided by the theory of \emph{jet spaces}. Classical field theory and, in particular, the calculus of variations have a nice geometric formulation within jet spaces. In this section we recall basic facts about them, and the main geometric constructions underlying field theory from both the Lagrangian and the Hamiltonian sides. At the same time, we set our (mathematical) notations.

We will consider PDEs imposed on sections of fiber bundles. The geometric portrait of a PDE is a submanifold in a jet space. Namely, let $\zeta : E \rightarrow M$ be a fiber bundle, $\dim M = m$, $\dim E = m + n$. We will often interpret $M$ as a space-time manifold. More generally it will be the \emph{manifold of independent variables}. We denote by $\Omega^i \rightarrow M$ the bundle of differential $i$-forms on $M$.  For the sake of simplicity, we assume $M$ to be oriented. This allow us to avoid the use of densities and to use differential forms, instead, as objects to be integrated over $M$. However, except for the orientation, $M$ will not carry any other extrinsic geometric structure, unless otherwise specified. We will often interpret $E$ as the target space of the fields, i.e.~a field is a section of $E$ (over $M$). Accordingly, fibers of $E$ are \emph{the manifolds of dependent variables}. The $k$-th order jet space encodes \emph{multiple, partial derivatives of dependent variables with respect to independent ones up to order} $k$ and can be defined as follows. Let $(x^i, u^\alpha)$ be bundle coordinates in $M$, i.e.~$(x^i)$ are local coordinates in $M$ and $(u^\alpha)$ are local fiber coordinates in $E$. A (local) section $\sigma$ of $E$ can be locally written as
\begin{equation}
\sigma : u^\alpha = f^\alpha (x^i)  \label{Luca1}
\end{equation}
for some smooth functions $(f^\alpha)$ of the $(x^i)$. We will consider multiple partial derivatives of the $(f^\alpha)$ with respect to the $(x^i)$. Our notations for partial derivatives are the following. Let $I = (i_1,\ldots, i_n)$ be an $n$-entry multiindex. We set
\[
\frac{\partial^{|I|} f^\alpha}{\partial x^I} := \frac{\partial^{i_1+\cdots+i_n} f^\alpha}{\partial x_1^{i_1} \cdots \partial x_n^{i_n}},\quad |I|{} := i_1 + \cdots + i_n.
\]

Two local sections $\sigma_1, \sigma_2$, with local description $\sigma_a : u^\alpha = f_a^\alpha (x^i)$, $a = 1,2$, are \emph{tangent up to order $k$ at a point $x \in M$} with local coordinates $(x^i)$ if
\[
\frac{\partial^{|I|} f_1^\alpha}{\partial x^I} (x^i) := \frac{\partial^{|I|} f_2^\alpha}{\partial x^I} (x^i), \quad |I|{}\leq k.
\]
Tangency up to order $k$ is a well defined equivalence relation. In particular, it is independent of the choice of coordinates. The equivalence class of section $\sigma$ is denoted by ${\mathsf j}^k \sigma (x)$ and it is called \emph{the $k$-th jet of $\sigma$ at $x$}. It contains a full, intrinsic information about derivatives of $\sigma$ at $x$ up to order $k$. For instance, the first jet of $\sigma$ at $x$ contains the same information as the tangent space to the image of $\sigma$ at $\sigma(x)$. The \emph{$k$-th jet space of sections of $E$} is the set ${\mathsf J}^k E$:
\[
{\mathsf J}^k E := \{ {\mathsf j}^k \sigma (x) : \sigma \text{ a local section of $E$ and $x \in M$}\}
\]
Clearly, ${\mathsf J}^0 E$ identifies with $E$, and ${\mathsf J}^1 E$ identifies with the set of $n$-dimensional tangent subspaces to $E$, transversal to fibers of $\pi$. Moreover, there are obvious projections $\zeta_k : {\mathsf J}^k E \rightarrow M$, ${\mathsf j}^k \sigma (x) \mapsto x$, and, $\zeta_{k,l} : {\mathsf J}^k E \rightarrow {\mathsf J}^l E$, ${\mathsf j}^k \sigma (x) \mapsto {\mathsf j}^l \sigma (x)$, $l \leq k$. In particular, $\zeta_{k,l}$ consists in ``dropping higher derivatives\textquotedblright . Clearly, $\zeta_k = \zeta_l \circ \zeta_{k,l} $, and $\zeta_{k,l} = \zeta_{p,l} \circ \zeta_{k,p}$, $l \leq p \leq k$.
The $k$-th jet space can be coordinatized as follows. Let $\mathcal{U}$ be a coordinate domain in $E$ and $(x^i , u^\alpha)$ bundle coordinates in it. There are \emph{jet coordinates} $(x^i,u^\alpha_I)$, $|I|{\leq k}$ on $\zeta_{k,0}^{-1}(\mathcal{U})$. Namely, pick ${\mathsf j}^k \sigma (x) \in \zeta_{k,0}^{-1}(\mathcal{U})$, and let $\sigma$ be locally given by (\ref{Luca1}). Then put
\[
x^i ( {\mathsf j}^k \sigma (x)) := x^i (x), \quad \text{and} \quad u^\alpha_I ({\mathsf j}^k \sigma (x)) := \frac{\partial^{|I|} f_1^\alpha}{\partial x^I} (x^i (x)).
\]
When equipped with jet coordinates, ${\mathsf J}^k E$ is a smooth manifold. Moreover, projections $\zeta_k$ and $\zeta_{k,l}$ are fiber bundles. For instance, a section of $\zeta_{1,0} : {\mathsf J}^1 E \rightarrow E$ can be understood as an Ehresmann connection in $E$, i.e.~an $n$-dimensional distribution on $E$, transversal to fibers of $\zeta$.

In the following, we will deal with various bundles and bundle maps. However, every manifold fibered over $M$ will be understood as a bundle over $M$ unless otherwise specified. For instance, we will understand the projection $\zeta_k$ and interpret ${\mathsf J}^k E$ as a bundle over $M$ without further comments. We will also consider jets of sections of various bundles. However, jets of section of $E$ will play a special role, and we denote simply by ${\mathsf J}^k$ the bundle ${\mathsf J}^k E$, if there is no risk of confusion.

A section $\sigma$ of $E$ can be prolonged to a section ${\mathsf j}^k \sigma : M \rightarrow {\mathsf J}^k$, $x \mapsto {\mathsf j}^k \sigma (x)$, called the \emph{$k$-th jet prolongation of $\sigma$}. If $\sigma$ is locally given by (\ref{Luca1}), then ${\mathsf j}^k \sigma$ is locally given by:
\[
{\mathsf j}^k \sigma : u^\alpha_I = \frac{\partial^{|I|} f^\alpha}{\partial x^I} (x^i),\quad |I|{}\leq k,
\]
and contains a full, intrinsic information about derivatives of $\sigma$ up to order $k$. Sections of ${\mathsf J}^k$ of the form ${\mathsf j}^k \sigma$ are sometimes called \emph{holonomic sections}.

The main geometric structure on ${\mathsf J}^k$, $k > 1$, consists in the following canonical embedding $\iota: {\mathsf J}^{k+1} \hookrightarrow {\mathsf J}^1 {\mathsf J}^{k}$, ${\mathsf j}^{k+1} \sigma (x) \mapsto {\mathsf j}^1 ({\mathsf j}^{k} \sigma ) (x)$. Denote by $(x^i,u^\alpha_I, u^\alpha_{I,i})$, $|I|{}\leq k$, jet coordinates in ${\mathsf J}^1{\mathsf J}^{k}$. Then $\iota$ is locally given by
\[
\iota (x^i,u^\alpha_I,u^\alpha_{I+j}) = (x^i, u^\alpha_I, u^\alpha_{I,j} = u^\alpha_{I+j}), \quad |I|{}\leq k,
\]
where, for $I = (i_1,\ldots,i_n)$ we denote by $I+j$ the multiindex $(i_1,\ldots,i_{j-1},i_j +1,i_{j+1},\ldots,i_n)$.
The embedding $\iota$ is able to \emph{detect holonomic sections of ${\mathsf J}^k$} in the following sense: \emph{a section $\Sigma$ of ${\mathsf J}^k$ is holonomic iff $j^1 \Sigma$ takes values in the image of $\iota$}. In the following, we will understand the map $\iota$ and interpret ${\mathsf J}^{k+1}$ as a (distinguished) submanifold of ${\mathsf J}^1 {\mathsf J}^k$. Elements of ${\mathsf J}^1 {\mathsf J}^k$ in ${\mathsf J}^{k+1}$ are sometimes called \emph{holonomic jets}.

There is another geometric structure on ${\mathsf J}^k$ which will be relevant for our purposes. Namely, the bundle $\zeta_{k,k-1}: {\mathsf J}^k \longrightarrow {\mathsf J}^{k-1}$ is an \emph{affine bundle} in a canonical way. The underlying vector bundle is the bundle $\vee^k {\mathsf T}^\ast M \otimes_{{\mathsf J}^{k-1}} {\mathsf V}  E$ whose fiber at a point ${\mathsf j}^{k-1} \sigma (x) \in {\mathsf J}^{k-1}$ is
\begin{equation}
\vee^k {\mathsf T}_x^\ast M \otimes_{\mathbb{R}} {\mathsf V} _{\sigma (x)} E. \label{Luca2}
\end{equation}
In (\ref{Luca2}) $\vee^k {\mathsf T}_x^\ast M $ is the $k$-th symmetric power of ${\mathsf T}_x^\ast M$, consisting of covariant, symmetric $k$-tensors on $M$ at $x$, and ${\mathsf V} _e E$ denotes the vertical tangent space to $E$ at $e$. In particular $\pi_{1,0}: {\mathsf J}^1 E \rightarrow E$ is an affine bundle modelled over ${\mathsf T}^\ast M \otimes_E {\mathsf V}  E$.

Jet spaces allow one to give an intrinsic definition of PDEs, i.e.~a definition manifestly independent of the choice of coordinates. Namely, a \emph{system of $k$-th order PDEs imposed on sections of the bundle $E$} is a (closed) submanifold $\mathcal{E}$ of ${\mathsf J}^k$. A \emph{solution} of a system of PDEs $\mathcal{E} \subset {\mathsf J}^k$ is a (local) section $\sigma$ of $E$ such that ${\mathsf j}^k \sigma$ takes values in $\mathcal{E}$. Locally, $\mathcal{E}$ is given by
\[
\mathcal{E} : F_a (x^i,u^\alpha_I)=0,\quad |I|{}\leq k,
\]
for some local functions $(F_a)$ on ${\mathsf J}^k$, and a section $\sigma$ of $E$, locally given by (\ref{Luca1}), is a solution iff
\[
F_a \left( x^i, \frac{\partial^{|I|} f^\alpha}{\partial x^I}(x^i) \right) =0.
\]
Thus, the analytic definition of systems of PDEs is recovered when using local coordinates.

There have been a lot of work about a possible geometric formulation of the Hamiltonian side of classical field theories (see \cite{Vi, Vi2, Vi3} for a recent proposal by the second author, see also references therein). Whatever the approach, affine duality plays a prominent role. Let us recall here basic facts about it. Let $N$ be a smooth manifold, $A \rightarrow N$ an affine bundle on it, and $\Lambda \rightarrow N$ a line bundle. In applications, $N$ will often be the total space of a bundle $N = E \rightarrow M$, $A$ will be the first jet bundle of $E$ and $\Lambda$ will be the bundle of $m$-forms on $M$ with values in functions on $E$, i.e.~the bundle $E \times_M \Omega^m$. For now, let us stick on the general case. Denote by $V$ the model vector bundle of $A$. Linear maps from fibers of $V$ to fibers of $\Lambda$ over the same point of $N$ form the vector bundle $A^\ast := V^\ast \otimes_N \Lambda$. Similarly, affine maps from fibers of $A$ to fibers of $\Lambda$ over the same point of $N$, form a vector bundle $A^\dag:=\mathsf{Aff}(A,\Lambda)$ over $N$. Moreover, there is a canonical projection $\mathrm{\ell}: A^\dag \rightarrow A^\ast$ which consists in \emph{taking the linear part}. It is easy to see that $\mathrm{\ell}$ is an affine bundle with $1$-dimensional fiber, and model vector bundle $A^\ast \times_N \Lambda \rightarrow A^\ast$.

We will need one more construction involving $A$ and $\Lambda$. Denote by ${\mathsf J}^1 \ell$ the space of first jets of sections of $\ell$. Sections of $\Lambda$ act on $A^\dag$ by vertical automorphisms in an obvious way. This action can be lifted to an action on ${\mathsf J}^1 \ell$ as follows. Let $\lambda$ be a section of $\Lambda$, and $H$ a generic section of $\ell$. Define the action of $\lambda$ on ${\mathsf j}^1 H (p)$, $p \in A^\dag$ as $\lambda . {\mathsf j}^1 H (p) := {\mathsf j}^1 (H + \lambda^\prime) (p)$ where $\lambda^\prime$ is a section of $A^\ast \times_N \Lambda \rightarrow A^\ast$, and, precisely, the pull-back of $\lambda$ via $A^\ast \rightarrow N$. The quotient of ${\mathsf J}^1 \ell$ with respect to the action of sections of $\Lambda$ is a smooth manifold denoted by ${\mathsf P} A^\dag$. Moreover, the induced projection ${\mathsf P} A^\dag \rightarrow A^\ast$ inherits from ${\mathsf J}^1 \ell \rightarrow A^\dag$ an affine bundle structure, with model vector bundle ${\mathsf V} ^\ast A^\ast \otimes_M \Lambda \rightarrow A^\ast$. Finally a section $H$ of $\ell$ can be ``differentiated''to get a section ${\mathrm d}^{\mathsf v}  H$ of ${\mathsf P} A^\dag \rightarrow A^\ast$, defined as the composition of ${\mathsf j}^1 H$ and the canonical projection ${\mathsf J}^1 \ell \rightarrow {\mathsf P} A^\dag$, according to the commutative diagram
\[
\begin{array}[c]{c}
\xymatrix@C=40pt{ {\mathsf J}^1 \ell \ar[d] \ar[r] & {\mathsf P} A^\dag \ar[d]  \\
                A^\dag \ar[r] & A^\ast \ar[ul]_-{{\mathsf j}^1 H} \ar@/^/[l]^-H \ar@/_/[u]_-{{\mathrm d}^{\mathsf v}  H}
}
\end{array}.
\]

The manifold ${\mathsf P} A^\dag$ is referred to as the \emph{affine phase space} \cite{GGU1,G}.
In the case when $E$ is a bundle over $M$, $A = {\mathsf J}^1 E$, and $\Lambda = E \times_M \Omega^m$, we have that $V = {\mathsf T}^\ast M \otimes_E {\mathsf V}  E$, hence $A^\ast = {\mathsf V} ^{\ast} E \otimes_E \Omega^{m-1}$, where ${\mathsf V} ^\ast E$ is the dual bundle to ${\mathsf V}  E$. In this case, we will also denote $A^{\ast}$ by $\mathcal{P}E$, or simply $\mathcal{P}$ if this does not lead to confusion, because it should be understood as the phase space of first order classical field theories defined on $E$ (see below). We will also denote $A^{\dag}$ by ${\mathsf J}^\dag E$, or just ${\mathsf J}^\dag$ if this does not lead to confusion. There is a \emph{tautological} vertical $1$-form $\vartheta_\mathcal{P}$ on $\mathcal{P}$ with values in $\Omega^{m-1}$, i.e.~a section of the bundle ${\mathsf V} ^{\ast} \mathcal{P} \otimes_\mathcal{P} \Omega^{m-1} \rightarrow \mathcal{P}$ defined as follows. Denote by $\pi : \mathcal{P} \rightarrow E$ the projection, and, for $p \in \mathcal{P}$ put
\[
(\vartheta_\mathcal{P})_p (\xi) := p (\pi_\ast (\xi)) \in \Omega^{m-1}, \quad \xi \in V_{p} E.
\]
Notice that $\vartheta_\mathcal{P}$ is a ``field theoretic version\textquotedblright of the \emph{Liouville $1$-form} on a cotangent bundle. The space ${\mathsf P} {\mathsf J}^\dag$ is a field theoretic version of a twice iterated cotangent bundle (see next section). See \cite{G} for an alternative description and more details.

Finally, recall that $i$-forms on $M$ with values in functions on $E$, i.e.~sections of the bundle $E \times_M \Omega^i$, can be ``differentiated along fibers of $E$'' to get vertical forms on $E$ with values in $\Omega^i$. Namely, there is an operator, the \emph{vertical differential} ${\mathrm d}^{\mathsf v} $ which takes a section $\sigma$ of $E \times_M \Omega^i$ to a section ${\mathrm d}^{\mathsf v}  \sigma$ of ${\mathsf V} ^\ast E \otimes_E \Omega^i$. In bundle coordinates, the vertical differential is given by
\[
{\mathrm d}^{\mathsf v}  \left( \sigma_{j_1 \cdots j_i} {\mathrm d} x^{j_1} \wedge \cdots \wedge {\mathrm d} x^{j_i} \right) = \frac{\partial  \sigma_{j_1 \cdots j_i}}{\partial u^\alpha}\, {\mathrm d}^{\mathsf v}  u^\alpha \otimes {\mathrm d} x^{j_1} \wedge \cdots \wedge {\mathrm d} x^{j_i},
\]
where the vertical differential ${\mathrm d}^{\mathsf v}  f$ of a function $f \in C^\infty (E)$ is just the restriction of ${\mathrm d} f$ to ${\mathsf V}  E$.

\section{First order field theory}\label{sec:3}

A Lagrangian field theory is specified by a variational principle of the kind (\ref{Luca3}). In the case of a $(k+1)$-st derivative theory, the Lagrangian density $L$ is a bundle map ${\mathsf J}^{k+1} \rightarrow \Omega^m$ covering the identiy. As we already remarked, ${\mathsf J}^{k+1}$ can be understood as a distinguished submanifold of ${\mathsf J}^1 {\mathsf J}^{k}$. Accordingly, a $(k+1)$-st derivative Lagrangian field theory on a bundle $E$ can be understood as a first derivative theory on ${\mathsf J}^k$ subjected to the (vakonomic) constraints ${\mathsf J}^{k+1} \subset {\mathsf J}^1{\mathsf J}^k$. This idea goes back to de Donder \cite{D}. Now, in mechanics, constraints in ${\mathsf T} M$ can be easily handled within the Tulczyjew triple approach (on an infinitesimal time interval, see Section \ref{sub:12}). Namely, a Lagrangian $L : C \rightarrow \mathbb{R}$ defined on a (constraint) submanifold $C$ of ${\mathsf T} M$ generates a Lagrangian submanifold $S_{C,L}$ in ${\mathsf T}^\ast {\mathsf T} M$ by putting
\[
S_{C,L} := \{ p_\xi \in {\mathsf T}^\ast {\mathsf T} M : \xi \in C \text{ and } \forall \delta x \in C,\  \langle p_\xi, \delta x \rangle= \langle {\mathrm d} L , \delta x \rangle \}.
\]
In its turn, $S_{C,L}$ determines a dynamics $\mathcal{D} := \alpha^{-1} (S_{C,L})$ in ${\mathsf T} {\mathsf T}^\ast M$. A similar construction works for constrained first derivative field theories (see \cite{G}), in particular, higher order field theories.

Thus, let us briefly recall the Tulczyjew triple for a first order field theory. Fields are sections of a bundle $\zeta: E\rightarrow M$ and a Lagrangian density is a bundle map ${\mathsf J}^1 \rightarrow\Omega^m$ covering the indentity. The details of the construction can be found in \cite{G} (see also \cite{dLMS}).

Precisely as for mechanics the Lagrangian side of the triple is based on variational calculus. The phase space for the theory is the total space of the bundle $\pi:\mathcal{P}\rightarrow E$, (see Section \ref{sec:2}). In the case with no sources the Lagrangian side of the Tulczyjew triple for first order classical field theories is then
\begin{equation*}
\begin{array}[c]{c}
\xymatrix@!C=1.5pc{  & {\mathsf J}^1\mathcal{P} \ar[rrr]^\alpha\ar[dl]\ar[ddr]|!{[dl];[drr]}\hole & & & \operatorname{Lag}E\ar[dl] \ar[ddr] & \\
 \mathcal{P}\ar@{=}[rrr] \ar[ddr] & & & \mathcal{P}\ar[ddr] & & \\
 & & {\mathsf J}^1 \ar@{=}[rrr]|!{[ur];[drr]}\hole  \ar[dl] & & & {\mathsf J}^1 \ar[dl] \\
 & E\ar@{=}[rrr]  & & & E &
}
\end{array},
\end{equation*}
where we denoted by $\operatorname{Lag} E$ the space ${\mathsf V} ^\ast{\mathsf J}^1 \otimes_{{\mathsf J}^1 E}\Omega^m$, and $\alpha$ is the field theoretic version of the Tulczyjew isomorphism $\alpha_M$ (see \cite{G} for its definition). In the following we shall  often use  $\operatorname{Lag}$ instead of $\operatorname{Lag} E$ if there is no risk of confusion. Both spaces, ${\mathsf J}^1\mathcal{P}$ and $\operatorname{Lag}$ are double affine-vector bundles \cite{GRU} with vector bundle structure over ${\mathsf J}^1 $  and affine bundle structure over $\mathcal{P}$. The map $\alpha$ is a double vector affine bundle morphism.
The phase equations determined by a Lagrangian density $L$ are the subset $\mathcal{D}$ of ${\mathsf J}^1\mathcal{P}$ given by
\begin{equation*}
\mathcal{D}=\alpha^{-1}\left({\mathrm d}^{\mathsf v}  L({\mathsf J}^1)\right).
\end{equation*}
The double bundle $\operatorname{Lag}$ on the right is endowed with a canonical vertical two-form  $\omega_{{\mathsf J}^1}$ with values in the line bundle $\Omega^m$. The form $\omega_{{\mathsf J}^1 }$ is fiber-wise symplectic (every fiber is in a sense a cotangent bundle). The double bundle ${\mathsf J}^1\mathcal{P}$ on the left is endowed with a canonical vertical two-form $\omega_{{\mathsf J}^1 \mathcal{P}}$ with values in $\Omega^m$ as well. Moreover, $\omega_{{\mathsf J}^1 \mathcal{P}}$ is fiber-wise presymplectic. Actually,
$$\omega_{{\mathsf J}^1\mathcal{P}}=\alpha^\ast\omega_{{\mathsf J}^1}.$$

The double bundle structure of $\operatorname{Lag}$ makes it easy to define the Legendre map $\lambda$:
\begin{equation}\label{lagr:3}
\lambda: \mathcal{P}\longrightarrow {\mathsf J}^1, \quad \lambda({\mathsf j}^1\sigma)= \xi({\mathrm d}^{\mathsf v}  L({\mathsf j}^1\sigma))
\end{equation}
where $\xi$ is the projection $\xi: Lag\rightarrow\mathcal{P}$. In coordinates
$$\lambda(x^i, u^\alpha, u^\alpha_j)=\left(x^i, u^\alpha, p^j_{\alpha}=\frac{\partial L}{\partial u^\alpha_j}\right).$$

We stress that, while the Tulczyjew morphism $\alpha_M:{\mathsf T}{\mathsf T}^\ast M\rightarrow{\mathsf T}^\ast{\mathsf T}^\ast M$ is an isomorphism, its field theoretic analogue $\alpha$ is not in general. One could reduce the space ${\mathsf J}^1\mathcal{P}$ to get a space isomorphic to the ``fiber-wise cotangent bundle\textquotedblright $Lag$, but then one would loose the obvious interpretation of the dynamics as a first order partial differential equation.

The Hamiltonian side of the Tuczyjew triple is
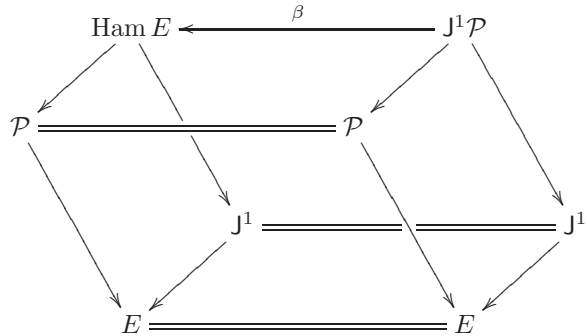
\begin{equation*}
\begin{array}[c]{c}
\xymatrix@!C=1.5pc{ & \operatorname{Ham}E\ar[dl]\ar[ddr]|!{[dl];[drr]}\hole & & & {\mathsf J}^1\mathcal{P} \ar[lll]_\beta \ar[dl] \ar[ddr] &  \\
\mathcal{P}\ar[ddr] & & & \mathcal{P} \ar@{=}[lll]\ar[ddr] & &  \\
    & & {\mathsf J}^1 \ar[dl] & & & {\mathsf J}^1  \ar@{=}[lll]|!{[dl];[ull]}\hole \ar[dl]  \\
 & E &  & & E \ar@{=}[lll] &
 }
\end{array}
\end{equation*}
where by $\operatorname{Ham} E$ we denoted the space ${\mathsf P}{\mathsf J}^\dag$, i.e.~the affine phase bundle for the affine dual bundle of ${\mathsf J}^1 \rightarrow E$ (see Section \ref{sec:2}). We shall often use $\operatorname{Ham}$ instead of $\operatorname{Ham} E$. The bundle $\operatorname{Ham}$ is a double affine-vector bundle with affine bundle structure over $\mathcal{P}$ and vector bundle structure over ${\mathsf J}^1$. The map $\beta$ is a double affine-vector bundle morphism
defined as the composition of the canonical isomorphism $\mathcal{R}_{{\mathsf J}^1}$ between $\operatorname{Lag}$ and $\operatorname{Ham}$ with $\alpha$
(see \cite{G} for the definition of $\mathcal{R}_{{\mathsf J}^1}$). Moreover $\operatorname{Ham}$ is endowed with a canonical vertical two-form $\omega_{{\mathsf J}^\dag}$ with values in $\Omega^m$. Since $\mathcal{R}_{{\mathsf J}^1}$ is an antisymplectomorphism we get that
\[
\beta^\ast \omega_{{\mathsf J}^\dag }=-\omega_{{\mathsf J}^1\mathcal{P}}.
\]

Phase equations $\mathcal{D}$ can be also generated by an \emph{affine generating object}, in the simplest case, a section $H$ of the bundle ${\mathsf J}^\dag\rightarrow \mathcal{P}$:
\begin{equation}\label{ham:4}
\mathcal{D}=\beta^{-1}({\mathrm d}^{\mathsf v}  H(\mathcal{P})).
\end{equation}
The next to the simplest case is when $\mathcal{D}= \alpha^{-1}({\mathrm d}^{\mathsf v}  L ({\mathsf J}^1))$ for a generic Lagrangian $L$. In this case, use symplectic relations techniques one obtains a \emph{generating family} of sections (of ${\mathsf J}^\dag\rightarrow \mathcal{P}$) parameterized by elements of ${\mathsf J}^1$:
\begin{equation*}
\tilde H: {\mathsf J}^1\times_E\mathcal{P}\rightarrow {\mathsf J}^\dag ,
\end{equation*}
which is equivalent to a family of form valued maps
\begin{equation*}
F_{\tilde H}:{\mathsf J}^1 \times_E {\mathsf J}^\dag \rightarrow \Omega^m,\qquad F_H({\mathsf j}^1\sigma(x), \varphi)=L({\mathsf j}^1\sigma(x))-\varphi({\mathsf j}^1\sigma(x)).
\end{equation*}
In some cases the above family reduces to a single generating section $H$. This is the field-theoretical version of Legendre transformation. Note, that a pair $({\mathsf j}^1\sigma(x), p)$ is critical for $\tilde{H}$ precisely if $p=\lambda({\mathsf j}^1\sigma(x))$.

Summarizing, the Tulczyjew triple for first order field theories on the bundle $E\rightarrow M$ is
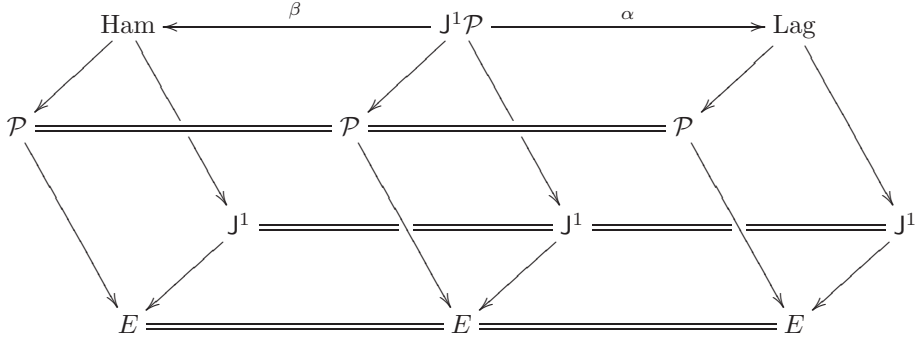
\begin{equation*}
\begin{array}[c]{c}
\xymatrix@!C=1.5pc{ & \operatorname{Ham} \ar[dl]\ar[ddr]|!{[dl];[drr]}\hole & & & {\mathsf J}^1\mathcal{P}\ar[lll]_\beta\ar[rrr]^\alpha\ar[dl]\ar[ddr]|!{[dl];[drr]}\hole & & & \operatorname{Lag}\ar[dl] \ar[ddr] & \\
\mathcal{P}\ar[ddr] & & & \mathcal{P}\ar@{=}[rrr]\ar@{=}[lll]\ar[ddr] & & & \mathcal{P}\ar[ddr] & & \\
    & & {\mathsf J}^1 \ar[dl] & & & {\mathsf J}^1 \ar@{=}[rrr]|!{[ur];[drr]}\hole \ar@{=}[lll]|!{[dl];[ull]}\hole \ar[dl] & & & {\mathsf J}^1 \ar[dl] \\
 & E &  & & E\ar@{=}[rrr]\ar@{=}[lll] & & & E &
}
\end{array}.
\end{equation*}
The right-hand-side is the Lagrangian one, the left-hand-side is the Hamiltonian one, the dynamics lives in the middle. Hamiltonian and Lagrangian spaces $\operatorname{Ham}$ and $\operatorname{Lag}$ are canonically isomorphic double affine-vector bundles equipped with (antisymplectomorphic) vertical symplectic forms with values in $\Omega^m$. The dynamics $\mathcal{D}$ is a submanifold of ${\mathsf J}^1\mathcal{P}$, i.e.~a first order partial differential equation on sections of the bundle $\mathcal{P}\rightarrow M$.

Finally, we remark that, exactly as in the opening of this section, a dynamics can be generated by a Lagrangian density $L$ even when $L$ is only defined on a ``constraint'' subbundle $C \subset {\mathsf J}^1$, $L : C \rightarrow \Omega^m$. Namely, first of all $L$ generates the ``Lagrangian'' submanifold $S_{C, L}$ in $\operatorname{Lag}$ given by
\begin{equation}\label{ur3}
S_{C, L} := \{\varphi_w \in Lag: w \in C \text{ and } \forall \delta w \in {\mathsf V}  C,\
\langle\varphi_w, \delta w\rangle=\langle{\mathrm d}^{\mathsf v}  L, \delta w\rangle\}.
\end{equation}
 Now put $\mathcal{D} := \alpha^{-1} (S_{C,L})$. It is easy to see that $\mathcal{D}$ is the ``correct phase dynamics'' of the Lagrangian field theory specified by the (vakonomic) constraints $C$ and the constrained Lagrangian $L$.

\section{Higher order field theory: The unreduced triple}\label{sec:4}

In this section we shall construct the Tulczyjew triple for field theories of order $(k+1)$ treated as constrained first order theories.
We interpret a $(k+1)$-st order Lagrangian $L:{\mathsf J}^{k+1}\rightarrow \Omega^{m}$ as a first order one defined on the submanifold ${\mathsf J}^{k+1}$ of holonomic jets in ${\mathsf J}^1 {\mathsf J}^k$. The price for going back to well-known structures is that we have to accept unphysical degrees of freedoms coming from the mathematical language.

In this approach we can repeat the construction of the Tulczyjew triple for first order theories replacing the bundle $\zeta: E\rightarrow M$ with the bundle $\zeta_k: {\mathsf J}^k\rightarrow M$ and using, in the simplest case, Lagrangians defined on ${\mathsf J}^{k+1}\subset{\mathsf J}^1{\mathsf J}^k$ as generating objects. Let us go through spaces and bundles that will appear in the triple.

Since Lagrangians are defined on a submanifold of ${\mathsf J}^1{\mathsf J}^k$  the Lagrangian space is
\begin{equation*}
\widetilde{\operatorname{Lag}}_k:= \operatorname{Lag}{\mathsf J}^k = {\mathsf V} ^\ast{\mathsf J}^1{\mathsf J}^k\otimes\Omega^m.
\end{equation*}
Using coordinates $(x^i, u^\alpha_I)$ in ${\mathsf J}^k$ as defined in Section \ref{sec:2} we can define the adapted coordinates $(x^i, u^\alpha_I, u^\beta_{J,j}, a_{\mu}^K, a_{\nu}^{K,k})$, with $|K| \leq k$, in $\widetilde{\operatorname{Lag}}_k$. Namely, a point in $\widetilde{\operatorname{Lag}}_k$ is an $\Omega^m$-valued vertical differential form on ${\mathsf J}^1 {\mathsf J}^k$ given by $(a_\mu^ K {\mathrm d}^{\mathsf v}  u^\alpha_K + a_\nu^{K,k} {\mathrm d}^{\mathsf v}  u^\nu_{K,k}) \otimes  \eta$, with $\eta := {\mathrm d} x^1 \wedge \cdots \wedge {\mathrm d} x^n$. The phase space is
\begin{equation*}
\mathcal{P}_k := \mathcal{P}{\mathsf J}^k = {\mathsf V} ^\ast{\mathsf J}^k\otimes\Omega^{m-1}
\end{equation*}
which is a vector bundle over ${\mathsf J}^k$. We can define a system of adapted coordinates $(x^i, u^\alpha_I, p^{I.j}_\beta)$, with $|I|\leq k$. Namely, a point in $\mathcal{P}_k$ is an $\Omega^{m-1}$-valued vertical form on ${\mathsf J}^k$ given by  $p^{I.i}_\alpha {\mathrm d}^{\mathsf v}  u^\alpha_I\otimes\eta_i$,where $\eta_i := (-1)^{i-1} {\mathrm d} x^1 \wedge \cdots \wedge \widehat{{\mathrm d} x^i} \wedge \cdots \wedge {\mathrm d} x^n$, and a hat ``$\widehat{(\cdot)}$\textquotedblright\ denotes omission.
The Tulczyjew morphism $\widetilde{\alpha}_k:= \alpha : {\mathsf J}^1\mathcal{P}_k\longrightarrow \widetilde{\operatorname{Lag}}_k$ is constructed exactly as in \cite{G}, replacing the bundle $E\rightarrow M$ with the bundle ${\mathsf J}^k\rightarrow M$. In adapted coordinates
\begin{equation*}
\widetilde{\alpha}_k(x^i, u^\alpha_I,\, p^{J.j}_\beta,\, u^\alpha_{I,i},\, p^{J.j}_{\beta}{}_{,i})=
\left(x^i, u^\alpha_I,\, u^\alpha_{I,i},\, a^J_\beta=\sum_{j=1}^m p^{J.j}_{\beta}{}_{,j},\, a^{J,j}_\beta=p^{J.j}_\beta\right).
\end{equation*}
The Lagrangian side of the Tulczyjew triple for field theories of order $(k+1)$ is then
\begin{equation}\label{ur:6}
\begin{array}[c]{c}
\xymatrix@!C=1.5pc{
& {\mathsf J}^1\mathcal{P}_k\ar[rrr]^{\widetilde{\alpha}_k}\ar[dl]\ar[ddr]|!{[dl];[drr]}\hole  & & &
 \widetilde{\operatorname{Lag}}_k \ar[dl]\ar[ddr] & \\
\mathcal{P}_k\ar@{=}[rrr]\ar[ddr] & & & \mathcal{P}_k \ar[ddr] & & \\
& & {\mathsf J}^1{\mathsf J}^k\ar[dl]\ar@{=}[rr]|!{[drr];[ur]}\hole & & & {\mathsf J}^1{\mathsf J}^{k} \ar[dl]\ar@{=}[lll]|!{[dl];[ull]}\hole  \\
& {\mathsf J}^k\ar@{=}[rrr] & & & {\mathsf J}^k &
}
\end{array}.
\end{equation}
Using Diagram (\ref{ur:6}) one can give the right answer, for instance, to the Question: ``What is the phase space dynamics of a field theory governed by a variational principle specified by a Lagrangian density $L : {\mathsf J}^{k+1} \rightarrow \Omega^m$?''
Indeed, the dynamics $\mathcal{D}_k$ should be a submanifold of ${\mathsf J}^1\mathcal{P}_k$, i.e.~a first order PDE on sections of the bundle $\mathcal{P}_k\rightarrow M$. Now, a dynamics can be generated from $L$ as discussed in the end of Section \ref{sec:3}. It is enough to put
\begin{equation*}
\mathcal{D}_k=\alpha_k^{-1}(S_{{\mathsf J}^{k+1}, L}).
\end{equation*}
(see (\ref{ur3})). A description of $\mathcal{D}_k$ in local coordinates shows that it is indeed the ``right answer'' to the above Question. Namely, in coordinates $(x^i, u^\alpha_I, u^\beta_{J,j}, a^I_\alpha, a^{J,j}_\beta)$ in $\widetilde{\operatorname{Lag}}_k$ the submanifold $S_{{\mathsf J}^{k+1}, L}$ reads
\begin{equation*}
\begin{array}{ll}\vspace{5pt}
u^\alpha_{J,j}=u^\alpha_{J+j} & |J|<k, \\ \vspace{5pt}
u^\alpha_{J+j}=u^\alpha_{I+i} & |J|=|I|=k, J+j=I+i, \\ \vspace{5pt}
a^I_\alpha+\delta^I_{J+i}a^{J,i}_\alpha=\frac{\partial L}{\partial u^\alpha_I} & |I|\leq k, \\ \vspace{5pt}
\delta^I_{J+i}a^{J,i}_\alpha=\frac{\partial L}{\partial u^\alpha_I} & |I|=k+1.
\end{array}
\end{equation*}
where symbol $\delta^I_{J+i}$ is a Kronecker delta-like symbol. It equals $1$ when multi-indices $I$ and $J+i$ coincide and it is $0$ otherwise.
Finally, in coordinates, the dynamics reads
\begin{equation*}
\begin{array}{ll}\vspace{5pt}
u^\alpha_{J,j}=u^\alpha_{J+j} & |J|<k, \\ \vspace{5pt}
u^\alpha_{J+j}=u^\alpha_{I+i} & |J|=|I|=k, J+j=I+i, \\ \vspace{5pt}
p^{I.j}_{\alpha}{}_{,j}+\delta^I_{J+i}p^{J.i}_\alpha=\frac{\partial L}{\partial u^\alpha_I} & |I|\leq k, \\ \vspace{5pt}
\delta^I_{J+i}p^{J.i}_\alpha=\frac{\partial L}{\partial u^\alpha_I} & |I|=k+1.
\end{array}
\end{equation*}

In first order field theory the double bundle structure of the Lagrangian side of the triple allowed us to construct the Legendre map (\ref{lagr:3}) that associates momenta to configurations. In the higher order case we do not have a map like this any more. This is because $S_{{\mathsf J}^{k+1}, L}$ is not the image of a section of the bundle $\widetilde{\operatorname{Lag}}_k\rightarrow{\mathsf J}^1{\mathsf J}^k$. Instead of a map we get only a relation. A point $p\in \mathcal{P}_k$ is in the relation $\lambda_k$ with $w\in{\mathsf J}^1{\mathsf J}^k$ if and only if $w$ is a holonomic jet and there exists a point in $S_{{\mathsf J}^{k+1}, L}$ that projects on both
$p$ and $w$. It is easy to see in coordinates that this means that
\begin{equation*}
\delta^I_{J+i}p^{J.i}_\alpha=\frac{\partial L}{\partial u^\alpha_I}\quad\text{for}\quad |I|=k+1.
\end{equation*}
One can say that only ``the highest order momenta'' are defined here. Since in the following we will often use relations we introduce a specific notation for them. Namely, a relation (as opposed to a plain map) will be indicated by a dotted line $\xymatrix@C=10pt{{} \ar@{.}[r] & {}}$. For instance
\begin{equation*}
\xymatrix{{\mathsf J}^1{\mathsf J}^k E\ar@{.}[r]^-{\lambda_k} & \mathcal{P}_k.}
\end{equation*}

Notice that $\mathcal{D}_k$ coincides exactly with the Euler-Lagrange-Hamilton equations determined by $L$ (see, e.g., \cite{Vi}, see also \cite{C...}). In their turn, as shown in \cite{Vi,C...} the Euler-Lagrange-Hamilton equations are essentially equivalent to the Euler-Lagrange equations, but still keep the nice feature of incorporating momenta. We conclude that diagram (\ref{ur:6}) contains a full information about the dynamics of a Lagrangian field theory (including the dynamics of momenta).

We now pass to the Hamiltonian side of the triple. The Hamiltonian space is $\widetilde{\operatorname{Ham}}_k := \operatorname{Ham}{\mathsf J}^k = {\mathsf P} {\mathsf J}^\dag{\mathsf J}^k$. Recall that the spaces $\widetilde{\operatorname{Ham}}_k$ and $\widetilde{\operatorname{Lag}}_k$ are canonically isomorphic double bundles and we can define the ``Hamiltonian Tulczyjew morphism'' $\widetilde{\beta}_k:{\mathsf J}^1\mathcal{P}_k\rightarrow\widetilde{\operatorname{Ham}}_k$ composing $\alpha_k$ with the canonical isomorphism $\widetilde{\operatorname{Lag}}_k \simeq \widetilde{\operatorname{Ham}}_k$. In coordinates the map $\widetilde{\beta}_k$ reads
\begin{equation*}
\widetilde{\beta}_k(x^i, u^\alpha_I, p^{J.j}_\beta, u^\alpha_{I,i}, p^{J.j}_{\beta}{}_{,i})=\left(x^i, u^\alpha_I, p^{J.j}_\beta, -\sum_j p^{J.j}_{\beta}{}_{,j}, u^\alpha_{I,i}\right).
\end{equation*}
The Hamiltonian side of the Tulczyjew triple for field theories of order $(k+1)$ is
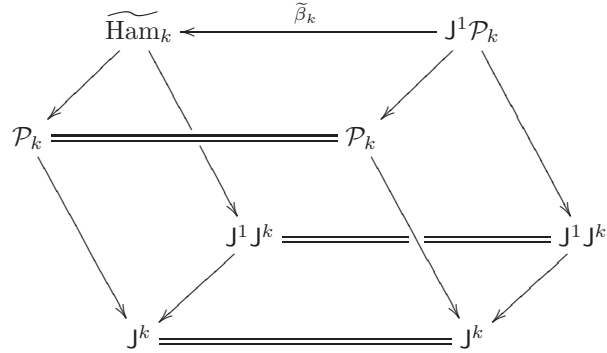
\begin{equation*}
\begin{array}[c]{c}
\xymatrix@!C=1.5pc{
&  \widetilde{\operatorname{Ham}}_k \ar[dl]\ar[ddr]|!{[dl];[drr]}\hole & & & {\mathsf J}^1\mathcal{P}_k\ar[lll]_{\widetilde{\beta}_k}\ar[dl]\ar[ddr] &  \\
\mathcal{P}_k \ar[ddr] & & &\mathcal{P}_k\ar@{=}[lll]\ar[ddr] & & \\
& & {\mathsf J}^1{\mathsf J}^{k} \ar[dl]\ar@{=}[rrr]|!{[drr];[ur]}\hole & & & {\mathsf J}^1{\mathsf J}^k\ar[dl]  \ar@{=}[ll]|!{[dl];[ull]}\hole\\
& {\mathsf J}^k\ar@{=}[rrr] & & & {\mathsf J}^k &
}
\end{array} .
\end{equation*}
Using the above diagram one can generate a dynamics $\mathcal{D}_k$ from an Hamiltonian generating object, in the simplest case a section $H$ of the bundle ${\mathsf J}^\dag{\mathsf J}^k\rightarrow \mathcal{P}_k$. In this case $\mathcal{D}_k=\beta_k^{-1}({\mathrm d}^{\mathsf v}  H(\mathcal{P}_k))$ exactly as (\ref{ham:4}) in first order theories.

Following the pattern of first order field theories we get $(k+1)$-st order version of Legendre transformation. For a generic Lagrangian $L$ the Hamiltonian generating object is actually a family of sections of the bundle ${\mathsf J}^\dag{\mathsf J}^k\rightarrow\mathcal{P}_k$ parameterized by elements of ${\mathsf J}^{k+1}$. It is easier to write the corresponding family of $\Omega^m$ valued maps
\begin{equation*}
F_{\tilde{H}}^k:{\mathsf J}^\dag{\mathsf J}^k\times_{{\mathsf J}^k}{\mathsf J}^{k+1}\rightarrow \Omega^m,\quad F_{\tilde{H}}^k(\varphi,{\mathsf j}^{k+1}\sigma(x))=L({\mathsf j}^{k+1}\sigma (x))-\varphi({\mathsf j}^1{\mathsf j}^{k}\sigma (x)) .
\end{equation*}
The pair $(\varphi,{\mathsf j}^{(k+1)}\sigma (x))$ is a critical point for the family $F_{\tilde{H}}^k$ if (in coordinates)
$$\frac{\partial L}{\partial u^\alpha_J}=\delta^J_{I+j}p^{I.j}_\alpha,\quad |J|=k+1,$$ i.e.~precisely when $\varphi$ projects on an element of $\mathcal{P}_k$ which is in the relation $\lambda_k$ with ${\mathsf j}^1{\mathsf j}^k\sigma (x)$.

The full version of the Tulczyjew triple for $(k+1)$-st derivative field theories is
\begin{equation}\label{ur:11}
\begin{array}[c]{c}
\xymatrix@!C=1pc{
 & \widetilde{\operatorname{Ham}}_k\ar[dl]\ar[ddr]|!{[dl];[drr]}\hole & & & {\mathsf J}^1\mathcal{P}_k\ar[lll]_{\widetilde{\beta}_k}\ar[rrr]^{\widetilde{\alpha}_k}\ar[dl]\ar[ddr]|!{[dl];[drr]}\hole & & &
 \widetilde{\operatorname{Lag}}_k \ar[dl]\ar[ddr] & \\
\mathcal{P}_k\ar[ddr] & & & \mathcal{P}_k\ar@{=}[rrr]\ar@{=}[lll]\ar[ddr] & & & \mathcal{P}_k \ar[ddr] & & \\
    & & {\mathsf J}^1{\mathsf J}^{k}\ar@{=}[rrr]|!{[ur];[drr]}\hole \ar[dl] & & & {\mathsf J}^1{\mathsf J}^k\ar[dl] \ar@{=}[ll]|!{[ull];[dl]}\hole \ar@{=}[rr]|!{[ur];[drr]}\hole& & & {\mathsf J}^1{\mathsf J}^{k} \ar[dl]\ar@{=}[lll]|!{[ull];[dl]}\hole \\
 & {\mathsf J}^k &  & & {\mathsf J}^k\ar@{=}[rrr]\ar@{=}[lll] & & & {\mathsf J}^k & }
 \end{array} .
\end{equation}

Summarizing, as usual the right-hand-side is the Lagrangian one, the left-hand-side is the Hamiltonian one, and the dynamics lives in the middle. Hamiltonian and Lagrangian spaces $\widetilde{\operatorname{Ham}}_k$ and $\widetilde{\operatorname{Lag}}_k$ are canonically isomorphic double affine-vector bundles equipped with (anti-symplectomorphic) vertical symplectic forms with values in $\Omega^m$. The dynamics $\mathcal{D}_k$ is a submanifold of ${\mathsf J}^1\mathcal{P}_k$ which can be generated either in Lagrangian or in Hamiltonian way.

We refer to diagram (\ref{ur:11}) as the ``unreduced Tulczyjew triple for $(k+1)$-st derivative field theories''. When using it for field theories depending on derivatives of the fields up to order $k+1$, we interpret ${\mathsf J}^{k+1}$ as a constraint subbundle in the ``configuration'' bundle ${\mathsf J}^1 {\mathsf J}^k$. However, the extra variables in ${\mathsf J}^1 {\mathsf J}^k$ are unphysical. In the next section we show that, starting from first principles, diagram
(\ref{ur:11}) can be actually ``reduced'' to a genuine (reduced) Tulczyjew triple where the unphysical degrees of freedom disappear.

\section{Higher order field theory: The reduced triple}\label{sec:5}

In the previous section we have constructed the Tulczyjew triple for $(k+1)$-st order field theories,
considering jets of order $(k+1)$ as holonomic first jets of sections of the bundle
${\mathsf J}^k\rightarrow M$. Following this point of view we could use well-developed first order theory for the price
of having non-physical degrees of freedom coming only from the mathematical language.

On the other hand Tulczyjew paradigms allow to construct a triple for virtually any theory, starting from first principles. Recall that the Lagrangian side of the Tulczyjew triple is directly obtained from variational calculus, while the Hamiltonian side can only be obtained after taking a suitable limit (see Section 1.2). In this section we follow Tulczyjew strategy to obtain a triple for field theories depending on higher derivatives of the fields. We call the result the ``reduced triple'' because it can also be obtained from the ``unreduced triple'' in previous section performing a suitable ``symplectic reduction''. More specifically, recall that the bundles $\widetilde{\operatorname{Ham}}_k$, ${\mathsf J}^1 \mathcal{P}_k$, and $\widetilde{\operatorname{Lag}}_k$ are equipped with canonical (line-bundle valued) vertical two-forms. All such two-forms are fiber-wise symplectic. Upon restricting them to holonomic jets they become presymplectic. The reduced triple can be then obtained by quotienting out the null-distribution of the restricted forms (see \cite{KM} for details), and it is free from unphysical degrees of freedom.

As usual, let the fields be sections of a fiber bundle $\zeta : E \rightarrow M$ over the ``space-time'' $M$. Let us first focus on fields propagating on a bounded domain $D$ of the space-time. We assume $D$ to have a smooth boundary $\partial D$. The configuration space $Q$ consists of sections $\sigma$ of $\zeta$ defined over $D$. We define processes, smooth functions and tangent vectors for $Q$ in a similar way as in Section \ref{sub:12}. Parametrized processes in $Q$ are vertical homotopies, i.e.~smooth maps $\chi:I\times D\rightarrow E$ where 1) $I$ is neighborhood of $0$ in ${\mathbb R}$, 2) for every $x \in D$, $s\mapsto \chi(s, x)$ is a vertical curve in $E$, and 3) for every $s\in I$, $x\mapsto \chi(s, x)$ is a section of $\zeta$. Functions on $Q$ are action functionals specified by $(k+1)$-st order Lagrangian densities $L:{\mathsf J}^{k+1}\rightarrow \Omega^m$ via the usual formula $S(\sigma)=\int_DL({\mathsf j}^{k+1}\sigma)$. Note that, as in mechanics, the composition of a function with a process is a smooth function $I\ni s\mapsto S(\chi(s, \cdot))\in {\mathbb R}$. Tangent vectors and covectors are obvious equivalence classes of processes and functions respectively. The pairing between vectors and covectors is given by the formula
\begin{equation*}
\langle {\mathrm d} S(\sigma), [\chi]\rangle=\int_D\frac{{\mathrm d}}{{\mathrm d} s}_{|s=0}L({\mathsf j}^k\chi(s,\cdot)),
\end{equation*}
where  ${\mathrm d} S(\sigma)$ is a tangent covector, the equivalence class of the pair $(S, \sigma)$, and $[\chi]$ is a tangent vector, the equivalence class of $\chi$. Adopting the Statics point of view reviewed in Section \ref{sec:1.1}, we interpret the $(k+1)$-st order field theory with Lagrangian $L$ as a regular system with cost function given by ${\mathrm d} S$ and constitutive set being $\mathcal{C}={\mathrm d} S(Q)$.

To find convenient representations of vectors and covectors we integrate by parts $k+1$ times and obtain
\begin{equation}\label{red:1}
\langle {\mathrm d} S(\sigma), [\chi]\rangle=\int_D \langle \mathcal{E}L ({\mathsf j}^{2k+2}\chi_{|s=0}), \delta\sigma\rangle+\int_{\partial D}\langle\mathcal{P}L({\mathsf j}^{2k+1}\chi_{|s=0}), \delta{\mathsf j}^k\sigma\rangle,
\end{equation}
where $\mathcal{E}L:{\mathsf J}^{2k+2}\rightarrow {\mathsf V} ^\ast E\otimes\Omega^m$ is the Euler-Lagrange morphism \cite{S}, $\mathcal{P}L: {\mathsf J}^{2k+1}\rightarrow {\mathsf V} ^\ast{\mathsf J}^k\otimes\Omega^{m-1}$ is a boundary term, $\delta\sigma$ denotes a vertical vector field on $E$ along $\sigma$ such that $\delta\sigma(x)$ is the tangent vector to curve $s\mapsto \chi(s,x)$ at $s=0$, and $\delta{\mathsf j}^k\sigma$ is the vertical vector field on ${\mathsf J}^k$ along ${\mathsf j}^k\sigma$ defined is a similar way.
It is easy to see that the tangent vector $[\chi]$ is equivalent to (i.e.~it contains the same information as) $\delta\sigma$. Similarly, the covector ${\mathrm d} S(\sigma)$ is equivalent to a pair $(f,p)$ where $f$ is a section of ${\mathsf V} ^\ast E\otimes\Omega^m\rightarrow M$ over $D$ and $p$ is a section of ${\mathsf V} ^\ast {\mathsf J}^k \otimes \Omega^{m-1}$ over $\partial D$. Using these convenient representations for tangent vectors and covectors, one sees that $\mathcal{C}$ is ``conveniently represented''by the following phase equations:
\[
\mathcal{D}=\{(f,p):\  f(x)=\mathcal{E}L({\mathsf j}^{2k+2}\sigma), \  p(x)=\mathcal{P}L({\mathsf j}^{2k+1}\sigma)\,\}.
\]
This means, in particular, that $f$ represents sources of the field, while ${\mathsf V} ^\ast{\mathsf J}^k\otimes\Omega^{m-1}$ should be understood as the phase space of the theory. Note that we have obtained the same phase space $\mathcal{P}_k$ as in the previous section.
It should be stressed, however, that the boundary term $\mathcal{P}L$ in (\ref{red:1}) is canonical only up to total differentials (see for instance \cite{Vi} and references therein). As a consequence, there are still ``non-physical degrees of freedom'' in $\mathcal{P}_k$. In principle, one could quotient them out at the price of loosing the nice interpretation of the dynamics as a submanifold in a jet space, i.e.~as a differential equation, which, on the other hand, is obviously desirable for many purposes. Therefore, we keep adopting $\mathcal{P}_k$ as the ``optimal'' phase space of the theory. In the following we shall consider only field theories without sources assuming $f=0$.

The Lagrangian side of the Tulczyjew triple is obtained, as in Section \ref{sub:12}, by passing to the new regime where the domain $D$ becomes infinitesimally small. It is easy to see that, in this infinitesimal limit, $Q$ becomes (a fiber in) ${\mathsf J}^{k+1}$. Accordingly, ${\mathsf T} Q$ becomes (a fiber in) ${\mathsf V}  J^{k+1}$, and ${\mathsf T}^\ast Q$ becomes (a fiber in) ${\mathsf V} ^\ast {\mathsf J}^{k+1} \otimes \Omega^m$. The latter space will be denoted by $\operatorname{Lag}_k E$, or simply $\operatorname{Lag}_k$ if this does not lead to confusion. It is naturally equipped with an obvious vertical symplectic form with values in $\Omega^m$. In the infinitesimal regime, a Lagrangian $L$ is interpreted as a potential for the cost function ${\mathrm d}^{\mathsf v}  L$, thus the constitutive set is $\mathcal{C}={\mathrm d}^{\mathsf v}  L({\mathsf J}^{k+1})\subset \operatorname{Lag}_k$. Now let us look at the correspondence between vectors and covectors and their convenient representations in the infinitesimal regime. Formula (\ref{red:1}), with the additional condition $f=0$, assumes the following form
\begin{equation}\label{red:2}
\langle {\mathrm d}^{\mathsf v}  L({\mathsf j}^{k+1}\sigma), \delta{\mathsf j}^{k+1}\sigma\rangle={\mathrm d}_M\langle p, \delta{\mathsf j}^k\sigma\rangle,
\end{equation}
where $p$ is a section of $\mathcal{P}_k$, and ${\mathrm d}_M$ is the \emph{total differential} \cite{G} (see also \cite{B,S} where ${\mathrm d}_M$ is referred to as the \emph{horizontal differential} and denoted differently). The right-hand side of (\ref{red:2}) defines a pairing between holonomic first jets of sections of $\mathcal{P}_k\rightarrow M$ and holonomic first jets of sections of ${\mathsf V} {\mathsf J}^k\rightarrow M$, where, by holonomic, we mean here ``projecting on holonomic jets in ${\mathsf J}^1{\mathsf J}^k$''. Let ${\mathsf j}^1 p(x) \in {\mathsf J}^1 \mathcal{P}_k$ and ${\mathsf j}^1 \delta \sigma (x) \in {\mathsf J}^1 {\mathsf V}  {\mathsf J}^k$ 1) project on holonomic jets in ${\mathsf J}^1 {\mathsf J}^k$ and, 2) project on the same jet ${\mathsf j}^k \sigma (x) \in {\mathsf J}^k$. Define the following pairing
\[
\langle\!\langle\, {\mathsf j}^1p(x),{\mathsf j}^{k+1}\delta\sigma(x)\,\rangle\!\rangle=
{\mathrm d}_M\langle\, p, \kappa_{k,1}({\mathsf j}^k\delta\sigma)\,\rangle(x),
\]
where $\kappa_{k,1}$ is the field theoretic version of the isomorphism $\kappa_M$ (\ref{mech:2})
\[
\kappa_{k,1}: {\mathsf J}^k {\mathsf V}  E\longrightarrow  {\mathsf V} {\mathsf J}^k, \quad  {\mathsf j}^k \delta \sigma (x) \longmapsto \delta {\mathsf j}^k \sigma (x).
\]
This shows that convenient representations of covectors are provided by points in ${\mathsf J}^1 \mathcal{P}_k$ projecting on holonomic jets in ${\mathsf J}^1 {\mathsf J}^k$, which we collectively denote by ${\mathsf J}^1_{hol} \mathcal{P}_k$. Similarly, convenient representations of vectors are given by points in ${\mathsf J}^1 {\mathsf V}  {\mathsf J}^k$ projecting on holonomic jets in ${\mathsf J}^1 {\mathsf J}^k$, or, which is the same, points in ${\mathsf J}^{k+1} {\mathsf V}  E$.

We now define a relation $\alpha_k$ generalizing $\alpha_M$. A covector $\psi\in \operatorname{Lag}_k$ is in the \emph{relation $\alpha_k$} with ${\mathsf j}^1p(x) \in{\mathsf J}^1\mathcal{P}_k$ if  1) ${\mathsf j}^1 p (x) \in {\mathsf J}^1_{hol} \mathcal{P}_k$, 2) $\psi$ and ${\mathsf j}^1 p (x)$ are over the same point of ${\mathsf J}^{k+1}$, and 3) for all ${\mathsf j}^{k+1}\delta\sigma \in {\mathsf J}^{k+1} {\mathsf V}  E$
\begin{equation*}
\langle\, \psi, \kappa_{k+1,1}^{-1}(\delta{\mathsf j}^{k+1}\sigma (x))\,\rangle=\langle\!\langle\, {\mathsf j}^1 p (x), {\mathsf j}^{k+1}\delta\sigma(x)\, \rangle\!\rangle.
\end{equation*}
In coordinates ${\mathsf j}^1p (x)$ is in the relation $\alpha_k$ with $\psi$ iff
\begin{equation*}
\begin{array}{ll}\vspace{5pt}
u^\alpha_I({\mathsf j}^1p (x))=u^\alpha_I(\psi) & |I|\leq k, \\ \vspace{5pt}
u^\alpha_{I,i}({\mathsf j}^1p (x))=u^\beta_{I+i}(\psi) & |I|\leq k, \\ \vspace{5pt}
p^{I.j}_{\alpha,j}({\mathsf j}^1p (x))+\delta^I_{J+i}p^{J.i}_\alpha({\mathsf j}^1p (x))=a_\alpha^I(\psi) & |I|\leq k, \\ \vspace{5pt}
\delta^I_{J+i}p^{J.i}_\alpha({\mathsf j}^1p (x))= a_\alpha^I(\psi)& |I|=k+1.
\end{array}
\end{equation*}
The relation $\alpha_k$
\begin{equation*}
\xymatrix@!C=3pc{{\mathsf J}^1\mathcal{P}_k\ar@{.}[r]^{\alpha_k} & \operatorname{Lag}_k}
\end{equation*}
is the main part of the Lagrangian side of the reduced Tuczyjew triple for $(k+1)$-st order field theories. Since the pairing $\langle\!\langle\cdot,\cdot\rangle\!\rangle$ is degenerate $\alpha_k$ is not an isomorphism and not even a map.

Before we present the full Lagrangian side of the reduced triple, let us examine the double bundle structure of $\operatorname{Lag}_k$. It is obviously a vector bundle over ${\mathsf J}^{k+1}$. The second bundle structure is an affine bundle. Recall that $\zeta_{k+1,k}:{\mathsf J}^{k+1}\rightarrow {\mathsf J}^k$ is an affine bundle with underlying vector bundle $\vee^{k+1}{\mathsf T}^\ast M\otimes_{{\mathsf J}^{k}}{\mathsf V}  E$ (see Section \ref{sec:2}). An element of $\operatorname{Lag}_k$ restricted to vectors tangent to the fibre of $\zeta_{k+1,k}$ acts as an element of $\vee^{k+1}{\mathsf T} M\otimes_{{\mathsf J}^{k}}{\mathsf V} ^\ast E\otimes_{{\mathsf J}^k}\Omega^m=: \mathcal{Q}_k$.  Accordingly there is a double vector-affine bundle
\begin{equation*}
\begin{array}[c]{c}
\xymatrix@!C=1.5pc{
& \operatorname{Lag}_k\ar[dr]^{\pi_{k}}\ar[dl]_{\xi_k} & \\
\mathcal{Q}_k\ar[dr] & & {\mathsf J}^{k+1}\ar[dl] \\
& {\mathsf J}^k &
}
\end{array}.
\end{equation*}
The Lagrangian side of the reduced Tulczyjew triple is the diagram
\begin{equation}\label{r:9}
\begin{array}[c]{c}
\xymatrix@!C=1.5pc{
 & {\mathsf J}^1\mathcal{P}_k\ar@{.}[rrr]^{\alpha_k}\ar[dl] \ar[ddr]|!{[ld];[rrd]}\hole & & &
 \operatorname{Lag}_k \ar[dl]\ar[ddr] & \\
\mathcal{P}_k\ar[rrr]\ar[ddr] & & & \mathcal{Q}_k \ar[ddr] & & \\
 & & {\mathsf J}^1{\mathsf J}^k\ar[dl] \ar@{-}[rr]|!{[ur];[drr]}\hole& & &\  {\mathsf J}^{k+1}\ar[dl] \ar@{^{(}->}[lll]|!{[ull];[dl]}\hole \\
 & {\mathsf J}^k\ar@{=}[rrr] & & & {\mathsf J}^k &
}
\end{array}.
\end{equation}
Both spaces ${\mathsf J}^1\mathcal{P}$ and $\operatorname{Lag}_k$ are double vector-affine bundles with vector bundle structures on the right and affine bundle structures on the left. The relation $\alpha_k$ is not a morphism of double vector affine bundles in a strict sense, but it is compatible with these structures. In particular $\alpha_k$ projects on the map $\mathcal{P}_k\rightarrow \mathcal{Q}_k$ which is a morphism of vector bundles over the identity on ${\mathsf J}^k$ defined as follows. The fibre of $\mathcal{P}_k$ over ${\mathsf j}^k\sigma(x)$ is the vector space $(\mathcal{P}_k)_{{\mathsf j}^k\sigma(x)}={\mathsf V} ^\ast_{{\mathsf j}^k \sigma(x)}{\mathsf J}^k\otimes\Omega^{m-1}_x$. Elements of this vector space act on ${\mathsf V} _{{\mathsf j}^k\sigma(x)}{\mathsf J}^k$ as one-forms with values in $\Omega^{m-1}_x$. The restriction of an element of $(\mathcal{P}_k)_{{\mathsf j}^k\sigma(x)}$ to vectors tangent to the fibre of $\zeta_{k,k-1}$ defines a projection from $(\mathcal{P}_k)_{{\mathsf j}^k\sigma(x)}$ to $\vee^k{\mathsf T}_xM\otimes{\mathsf V} ^\ast_{\sigma(x)}E\otimes\Omega^{m-1}_x$. On the other hand $(\mathcal{Q}_k)_{{\mathsf j}^k\sigma(x)}=\vee^{k+1}{\mathsf T}_x M\otimes{\mathsf V} ^\ast_{\sigma(x)}E\otimes\Omega^m_x$, and there is a ``symmetrization map''
$$\vee^k{\mathsf T}_xM\otimes{\mathsf V} _\sigma(x)E\otimes\Omega^{m-1}_x\longrightarrow (\mathcal{Q}_k)_{{\mathsf j}^k\sigma(x)}$$
that in coordinates reads
$$p^{I.i}_\alpha\partial_{x^{I}}\otimes{\mathrm d}^{\mathsf v} u^\alpha\otimes\eta_i\longmapsto
(\delta_{I+i}^J p^{I.i}_\alpha) \partial_{x^{J}}\otimes{\mathrm d}^{\mathsf v} u^\alpha\otimes\eta,$$
where, for $I = i_1 \cdots i_k$, we put $\partial_{x^I} = \partial_{x^{i_1}} \vee \cdots \vee \partial_{x^{i_k}}$.
Composing, we get the map $\mathcal{P}_k\rightarrow \mathcal{Q}_k$ in (\ref{r:9}).

Finally let us observe that, since ${\mathsf J}^{k+1}\subset {\mathsf J}^1 {\mathsf J}^k$, there is an obvious cotangent relation
\begin{equation*}
\xymatrix@!C=3pc{\widetilde{\operatorname{Lag}}_k\ar@{.}[r]^{\rho_k} & \operatorname{Lag}_k.}
\end{equation*}
Namely, recall that $\widetilde{\operatorname{Lag}}_k = {\mathsf V}^\ast {\mathsf J}^1 {\mathsf J}^k \otimes \Omega^m$, and $\operatorname{Lag}_k = {\mathsf V}^\ast {\mathsf J}^{k+1} \otimes \Omega^m$. In particular, they fit into the commutative diagram
\[
\xymatrix{ \operatorname{Lag}_k \ar[d] & i^\ast (\widetilde{\operatorname{Lag}}_k ) \ar[l]_-{\rho} \ar[r] \ar[dl] & \widetilde{\operatorname{Lag}}_k \ar[d] \\
{\mathsf J}^{k+1}\  \ar@{^{(}->}[rr]^-{i} & & {\mathsf J}^1{\mathsf J}^k
}
\]
where $i$ is the inclusion, and $\rho$ is the restriction to ${\mathsf J}^{k+1}$ of vertical covectors on ${\mathsf J}^1{\mathsf J}^k$. The composition $\operatorname{Lag}_k \leftarrow i^\ast (\widetilde{\operatorname{Lag}}_k ) \rightarrow \widetilde{\operatorname{Lag}}_k $ is, by definition $\rho_k$. Now, diagram
\begin{equation*}
\begin{array}[c]{c}
\xymatrix@!C=1.5pc{\widetilde{\operatorname{Lag}}_k\ar@{.}[rr]^{\rho_k} & & \operatorname{Lag}_k \\
 & {\mathsf J}^1\mathcal{P}_k\ar[ul]^{\tilde\alpha_k}\ar@{.}[ur]_{\alpha_k} & }
 \end{array}
\end{equation*}
commutes (in a \emph{relation-theoretic} sense). As a consequence, the phase equations in $(k+1)$-st order theory obtained by means of the Lagrangian side of the reduced and
unreduced triples are the same, i.e.
\[
\widetilde{\alpha}_k^{-1}(S_{{\mathsf J}^{k+1},L})=\alpha_k({\mathrm d}^{\mathsf v}  L({\mathsf J}^{k+1})).
\]

In the reduced triple we can also find the Legendre relation $\lambda_k$ expressed as the composition of ${\mathrm d}^{\mathsf v}  L$, $\alpha_k$ and the projection ${\mathsf j}^1\mathcal{P}_k\rightarrow \mathcal{P}_k$, as illustrated in the following diagram
\begin{equation*}
\begin{array}[c]{c}
\xymatrix@!C=1.5pc{
 & {\mathsf J}^1\mathcal{P}_k\ar@{.}[rrr]^{\alpha_k}\ar[dl]\ar[ddr]|!{[ld];[rrd]}\hole|!{[ld];[rrrrdd]}\hole & & & \operatorname{Lag}_k \ar[dl]\ar[ddr] & \\
\mathcal{P}_k\ar[rrr]\ar[ddr] & & & \mathcal{Q}_k \ar[ddr] & & \\
 & & {\mathsf J}^1{\mathsf J}^k\ar[dl] \ar@{-}[rr]|!{[ur];[drr]}\hole& & &\  {\mathsf J}^{k+1}\ar[dl]\ar@{^{(}->}[lll]|!{[ull];[dl]}\hole \ar@/_1pc/[uul]_{{\mathrm d}^{\mathsf v}  L}\ar@{.}[ulllll]_{\lambda_k}|!{[ull];[dl]}\hole \\
 & {\mathsf J}^k\ar[rrr] & & & {\mathsf J}^k &
}
\end{array}.
\end{equation*}
The structure of the relation $\lambda_k$ is clear: using the double bundle $\operatorname{Lag}_k$
we can construct a map $\ell_k:{\mathsf J}^{k+1}\rightarrow \mathcal{Q}_k$ as a composition
\begin{equation}\label{r:11}
\begin{array}[c]{c}
\xymatrix@!C=1.5pc{
& \operatorname{Lag}_k\ar[dr]\ar[dl]_{\xi_k} & \\
Q_k\ar[dr] & & J^{k+1}\ar[dl]\ar@/_1pc/[ul]_{{\mathrm d}^{\mathsf v}  L}\ar[ll]_{\ell_k} \\
& J^k &
}
\end{array} , \qquad \qquad    \ell_l=\xi_k\circ{\mathrm d}^{\mathsf v}  L.
\end{equation}
The element ${\mathsf j}^{k+1}\sigma(x)$ is in the relation $\lambda_k$ with $p\in\mathcal{P}_k$ if $p$ projects
on $\ell_k({\mathsf j}^{k+1}\sigma(x))\in \mathcal{Q}_k$. This means that $\lambda_k({\mathsf j}^{k+1}\sigma(x))$ contains the whole inverse image of $\ell_k({\mathsf j}^{k+1}\sigma(x))$ with respect to $\mathcal{P}_k\rightarrow \mathcal{Q}_k$.

\medskip

Now we pass to the Hamiltonian side of the reduced triple, which is based on the affine bundle structure of $\zeta_{k+1,k}:{\mathsf J}^{k+1}\rightarrow {\mathsf J}^k$. The space of affine maps on fibres of $\zeta_{k+1,k}$ with values in the appropriate fibre of $\Omega^m$ will be denoted, for simplicity, by $\mathcal{K}_k$. Note that there is a canonical projection $\mathcal{K}_k\rightarrow \mathcal{Q}_k$ consisting of taking the linear part of an affine map. The bundle $\mathcal{K}_k\rightarrow \mathcal{Q}_k$ is an affine bundle with one dimensional fibre. The underlying vector bundle is $\mathcal{Q}_k\times_M\Omega^m\rightarrow \mathcal{Q}_k$. The affine phase bundle ${\mathsf P}\mathcal{K}_k$ (see Section \ref{sec:2}) is the Hamiltonian space for the Hamiltonian side of the reduced triple for $(k+1)$-st order field theories. It will be denoted by $\operatorname{Ham}_k E$, or simply $\operatorname{Ham}_k$.

The Hamiltonian space $\operatorname{Ham}_k$ is a double vector affine bundle with affine bundle structure over $\mathcal{P}_k$ and vector bundle structure over ${\mathsf J}^{k+1}$. Moreover, it is easy to see (along very similar lines as in \cite{G}), that $\operatorname{Ham}_k$ is naturally equipped with a vertical symplectic form with values in $\Omega_m$. Finally, $\operatorname{Ham}_k$ is canonically isomorphic to $\operatorname{Lag}_k$, and the canonical isomorphism $\mathcal{R}_k=\mathcal{R}_{{\mathsf J}^{k+1}}$ is an anti-symplectomorphism with respect to the ``canonical structures'' on $\operatorname{Lag}_k$ and $\operatorname{Ham}_k$. In $\operatorname{Ham}_k$ we shall use the coordinates $(x^i, u^\alpha_I, f^J_\beta, \xi^I_\alpha, \zeta^\alpha_J)$ where $|I|\leq k$ and $|J|=k+1$. In coordinates $\upsilon\in \operatorname{Ham}_k$ is identified with an element of ${\mathsf V} ^\ast\mathcal{Q}_k\otimes\Omega^m$ and $\upsilon=\xi^I_\alpha{\mathrm d}^{\mathsf v} u^I_\alpha\otimes\eta+\zeta^\beta_J{\mathrm d}^{\mathsf v} f^J_\beta\otimes \eta$. In coordinates, the canonical isomorphism $\mathcal{R}_k$ reads
\begin{equation*}
\mathcal{R}_k(x^i, u^\alpha_I, u^\beta_J, a^I_\alpha, a^J_\beta)=(x^i, u^\alpha_I, f^J_\beta=a^J_\beta, \xi^I_\alpha=a^I_\alpha, \zeta^\beta_J=-u^\beta_J), \hspace{5pt} |I|\leq k,\  |J|=k+1.
\end{equation*}
The composition of $\mathcal{R}_k$ and $\alpha_k$ is the relation $\beta_k$
\begin{equation*}
\beta_k=\mathcal{R}_k\circ\alpha_k
\end{equation*}
which is the Hamiltonian side of the reduced Tulczyjew triple
\begin{equation*}\label{r:12}
\begin{array}[c]{c}
\xymatrix@!C=1.5pc{
 & \operatorname{Ham}_k\ar[dl]\ar[ddr]|!{[rrd];[dl]}\hole & & & {\mathsf J}^1\mathcal{P}_k\ar@{.}[lll]_{\beta_k}\ar[dl]\ar[ddr] & \\
\mathcal{Q}_k\ar[ddr] & & & \mathcal{P}_k\ar[lll]\ar[ddr] & &  \\
    & & {\mathsf J}^{k+1}\ \ar@{^{(}->}[rrr]|!{[ur];[drr]}\hole\ar[dl] & & & {\mathsf J}^1{\mathsf J}^k\ar[dl] \ar@{-}[ll]|!{[ull];[dl]}\hole \\
 & {\mathsf J}^k &  & & {\mathsf J}^k\ar@{=}[lll] &
}
\end{array} .
\end{equation*}
In coordinates $\upsilon\in \operatorname{Ham}_k$ is in the relation $\beta_k$ with ${\mathsf j}^1 p(x)\in{\mathsf J}^1\mathcal{P}_k$ if
\begin{equation*}
\begin{array}{ll}\vspace{5pt}
u^\alpha_I({\mathsf j}^1p (x))=u^\alpha_I(\upsilon) & |I|\leq k, \\ \vspace{5pt}
u^\alpha_{I,i}({\mathsf j}^1p (x))=u^\beta_{I+i}(\upsilon) & |I|< k, \\ \vspace{5pt}
u^\alpha_{I,i}({\mathsf j}^1p (x))=-\zeta^\beta_{I+i}(\upsilon) & |I|=k, \\ \vspace{5pt}
p^{I.j}_{\alpha,j}({\mathsf j}^1p (x))+\delta^I_{J+i}p^{J.i}_\alpha({\mathsf j}^1p (x))=\xi_\alpha^I(\upsilon) & |I|\leq k, \\ \vspace{5pt}
\delta^I_{J+i}p^{J.i}_\alpha({\mathsf j}^1p (x))= f_\alpha^I(\upsilon)& |I|=k+1.
\end{array}
\end{equation*}
Similarly as above, one sees that the phase field equations $\mathcal{D}_k = \alpha_k ({\mathrm d}^{\mathsf v}  L ({\mathsf J}^{k+1}))$ generated by a generic Lagrangian, are also generated, on the Hamiltonian side by a family of sections of the bundle $\mathcal{K}_k\rightarrow \mathcal{Q}_k$ parametrerized by point in ${\mathsf J}^{k+1}$. In its turn, this family is equivalent to the family of $\Omega^m$ valued maps
\begin{equation*}
F: \mathcal{K}_k\times_{{\mathsf J}^k} {\mathsf J}^{k+1} \longrightarrow \Omega^m, \quad F(\varphi, {\mathsf j}^{k+1}\sigma (x)) = \varphi({\mathsf j}^k\sigma (x))-L({\mathsf j}^{k+1}\sigma (x)).
\end{equation*}

The complete reduced Tulczyjew triple for theories of order $(k+1)$ is
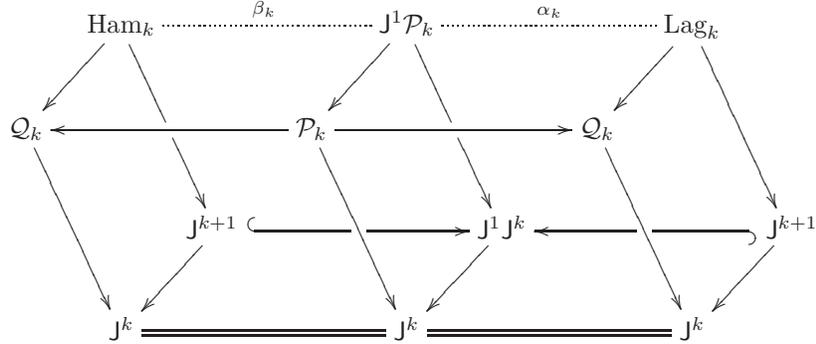
\begin{equation*}
\begin{array}[c]{c}
\xymatrix@!C=1pc{
 & \operatorname{Ham}_k\ar[dl]\ar[ddr]|!{[rrd];[dl]}\hole & & & {\mathsf J}^1\mathcal{P}_k\ar@{.}[lll]_{\beta_k}\ar@{.}[rrr]^{\alpha_k}\ar[dl]\ar[ddr]|!{[rrd];[dl]}\hole & & &
 \operatorname{Lag}_k \ar[dl]\ar[ddr] & \\
\mathcal{Q}_k\ar[ddr] & & & \mathcal{P}_k\ar[rrr]\ar[lll]\ar[ddr] & & & \mathcal{Q}_k \ar[ddr] & & \\
    & & {\mathsf J}^{k+1}\ \ar@{^{(}->}[rrr]|!{[ur];[drr]}\hole \ar[dl] & & & {\mathsf J}^1{\mathsf J}^k\ar[dl] \ar@{-}[rr]|!{[ur];[drr]}\hole \ar@{-}[ll]|!{[ull];[dl]}\hole& & &\  {\mathsf J}^{k+1}\ar[dl]\ar@{^{(}->}[lll]|!{[ull];[dl]}\hole \\
 & {\mathsf J}^k &  & & {\mathsf J}^k\ar@{=}[rrr]\ar@{=}[lll] & & & {\mathsf J}^k &
}
\end{array} .
\end{equation*}

\section{On the relation between different order triples}

Tulczyjew triples (\ref{ur:11}) and (\ref{r:11}) capture most of the relevant geometric structures underlying classical field theories of order $k+1$ defined on the bundle $E$. For instance, they prescribe how to produce the dynamics from different kind of generating objects (e.g., a Lagrangian density defined on a constraint submanifold) ``depending on derivatives of the fields up to order $k+1$''. Notice that when a ``generating object depends on derivatives up to order $k+1$'', one may safely state that ``it also depends on derivatives up to order $l+1$'' for all $l \geq k$. For instance, any $(k+1)$-st order Lagrangian density is an $(l+1)$-st order Lagrangian density as well, for every $l \geq k$. In other words, generating objects can be pull-backed to higher order jet bundles. In this section we want to give a precise mathematical meaning to this claim. In particular we shall discuss the relationship between different order Tulczyjew triples. In order to do this, it is convenient to start from the relationship between the triples of first order field theories defined on two different bundles connected by a bundle morphism.

Let $G,F$ be bundles over the same manifold $M$, and let $\Phi : G \rightarrow F$ be a bundle morphism over the identity of $M$, i.e.~$\Phi$ is a smooth map such that diagram
\begin{equation} \label{Luca4}
\begin{array}[c]{c}
\xymatrix{      G \ar[d] \ar[r]^-{\Phi} & F \ar[d] \\
                      M \ar@{=}[r] & M}
                      \end{array}
\end{equation}
commutes. Recall that diagram (\ref{Luca4}) can be prolonged to a diagram
\begin{equation}\label{Luca5}
\begin{array}[c]{c}
\xymatrix{     {\mathsf J}^1 G \ar[d] \ar[r]^-{{\mathsf j}^1 \Phi} & {\mathsf J}^1 F \ar[d] \\
                      G \ar[d] \ar[r]^-{\Phi} & F \ar[d] \\
                      M \ar@{=}[r] & M}
                      \end{array} ,
\end{equation}
where the map ${\mathsf j}^1 \Phi$ is defined as ${\mathsf j}^1\Phi({\mathsf j}^1 \sigma (x)) := {\mathsf j}^1 (\Phi \circ \sigma) (x)$ and can be characterized as the unique bundle map making diagram (\ref{Luca5}) commutative and mapping holonomic sections to holonomic sections. In particular, ${\mathsf j}^1 \Phi$ is a morphism of affine bundles over $\Phi$. Its linear part ${\mathsf T}^\ast M \otimes_G {\mathsf V}  G \rightarrow {\mathsf T}^\ast M \otimes_F {\mathsf V}  F$ is nothing but the well defined restriction ${\mathsf V}  \Phi : {\mathsf V}  G \rightarrow {\mathsf V}  F$ of the tangent map ${\mathsf T} \Phi : {\mathsf T} G \rightarrow {\mathsf T} F$ to $\Phi$ tensor the identity of ${\mathsf T}^\ast M$.

Let us now discuss the relationship between $\mathcal{P}G$ and $\mathcal{P} F$. Sections of the bundle $\mathcal{P}F \rightarrow F$ are $\Omega^{m-1}$-valued, vertical $1$-forms on $F$. As such, they can be pulled-back to sections of $\mathcal{P}G \rightarrow G$ via the bundle morphism $\Phi$, but there is no natural bundle morphism $\mathcal{P}F \rightarrow \mathcal{P}G$. However, there is a natural bundle relation between $\mathcal{P}G$ and $\mathcal{P}F$ covering $\Phi$. Namely, The pull back bundle $X := G \times_F \mathcal{P} F$ maps to $\mathcal{P}G$ as follows: $(e, \omega) \mapsto ({\mathsf V} _e^\ast \Phi)(\omega)$, $(e, \omega) \in G \times_F \mathcal{P} F$. Accordingly, there is a commutative diagram
\begin{equation} \label{Luca6}
\begin{array}[c]{c}
\xymatrix@C=15pt{ \mathcal{P}G \ar[d] & X \ar[r] \ar[l] \ar[dl]  & \mathcal{P}F \ar[d]\\
G \ar[d] \ar[rr]^-{\Phi} & & F \ar[d] \\
                      M \ar@{=}[rr] & & M}
                      \end{array} .
\end{equation}
Notice that the relation $\xymatrix@C=10pt{\mathcal{P}G \ar@{.}[r] & \mathcal{P}F}$ obtained in this way is just a bundle-theoretic version of the standard cotangent lift of a smooth map. Now, all arrows originating from $X$ are bundle morphisms over the identity of $M$. Thus, diagram (\ref{Luca6}) prolongs to a diagram
\begin{equation*}
\begin{array}[c]{c}
\xymatrix@C=15pt{ {\mathsf J}^1\mathcal{P}G \ar[d] & {\mathsf J}^1 X \ar[r] \ar[l] \ar[d] \ar[dl]  & {\mathsf J}^1 \mathcal{P}F \ar[d]\\
\mathcal{P}G \ar[d] & X \ar[r] \ar[l] \ar[dl]  & \mathcal{P}F \ar[d]\\
G \ar[d] \ar[rr]^-{\Phi} & & F \ar[d] \\
                      M \ar@{=}[rr] & & M}
                      \end{array} .
\end{equation*}
In a very similar way, one can construct diagrams
\begin{equation*}
\begin{array}[c]{c}
\xymatrix@C=15pt{ \operatorname{Lag}G \ar[d] & Y \ar[r] \ar[l]  \ar[dl]  & \operatorname{Lag} F \ar[d]\\
{\mathsf J}^1 G \ar[d]  \ar[rr]^-{{\mathsf j}^1 \Phi} &   & {\mathsf J}^1 F \ar[d]\\
G \ar[d] \ar[rr]^-{\Phi} & & F \ar[d] \\
                      M \ar@{=}[rr] & & M}
                      \end{array} \quad \text{and} \quad
                      \begin{array}[c]{c}
\xymatrix@C=15pt{ \operatorname{Ham}G \ar[d] & Z \ar[r] \ar[l]  \ar[dl]  & \operatorname{Ham} F \ar[d]\\
{\mathsf J}^1 G \ar[d]  \ar[rr]^-{{\mathsf j}^1 \Phi} &   & {\mathsf J}^1 F \ar[d]\\
G \ar[d] \ar[rr]^-{\Phi} & & F \ar[d] \\
                      M \ar@{=}[rr] & & M}
                      \end{array},
\end{equation*}
where $Y := \operatorname{Lag} F \times_{{\mathsf J}^1 F} {\mathsf J}^1 G$, and $Z := \operatorname{Ham} F \times_{{\mathsf J}^1 F} {\mathsf J}^1 G$. In particular there are natural relations $\xymatrix@C=10pt{{\mathsf J}^1 \mathcal{P} G \ar@{.}[r] & {\mathsf J}^1 \mathcal{P} F}$, $\xymatrix@C=10pt{\operatorname{Lag}G \ar@{.}[r] & \operatorname{Lag} F}$ and $\xymatrix@C=10pt{\operatorname{Ham}G \ar@{.}[r] & \operatorname{Ham} F}$. All of them do actually preserve the canonical structures. Details are left to the reader.

It is easy to see that diagram
\begin{equation}\label{Luca12}
\begin{array}[c]{c}
\xymatrix@C=3pt{       &        &       &    &                        & \operatorname{Ham} F \ar[dr]|!{[rr];[dlll]}\hole|!{[rrrr];[dl]}\hole \ar@{.}[dlllll]& & {\mathsf J}^1 \mathcal{P}F \ar[rr] \ar[ll] \ar[dr]|!{[rr];[dlll]}\hole \ar[dl]|!{[rr];[dlll]}\hole \ar@{.}[dlllll] & & \operatorname{Lag} F \ar[dl] \ar@{.}[dlllll]\\
                 \operatorname{Ham} G \ar[dr]&  & {\mathsf J}^1 \mathcal{P}G \ar[rr] \ar[ll] \ar[dr] \ar[dl]& & \operatorname{Lag} G  \ar[dl]&  &  \mathcal{P}F \ar[dr]|!{[rr];[dlll]}\hole \ar@{.}[dlllll]|!{[ll];[dlll]}\hole|!{[llll];[dlll]}\hole & & {\mathsf J}^1 F \ar@{<-}[dlllll]^-{{\mathsf j}^1 \Phi} \ar[dl]& \\
 &  \mathcal{P}G \ar[dr] & & {\mathsf J}^1 G \ar[dl]& & &  & F \ar@{<-}[dlllll]^-{\Phi}& & \\
 & &  G &  & & & & & &
}
\end{array}
\end{equation}
commutes. In this sense morphism $\Phi$ lifts to a relation between Tulczyjew triples. This clarifies the relationship between Tulczyjew triples of $G$ and $F$. In the special case when $G = {\mathsf J}^l$, $F = {\mathsf J}^k$, and $\Phi = \zeta_{l,k}$, $l \geq k$, we get a relation between the $(k+1)$-st order, and the $(l+1)$-st order, unreduced Tulczyjew triples of $E$.

Now, we look at the relationship between the dynamics generated by generating objects which are ``related through diagram (\ref{Luca12})''. For simplicity, instead of considering the most general situation we will only consider one case among the most interesting ones: a Lagrangian density $L : C \rightarrow \Omega^m$ is assigned on a subbundle $C \subset {\mathsf J}^1 F$ of ${\mathsf J}^1 F \rightarrow F$. Recall that $L$ generates a dynamics $\mathcal{D}_L = \alpha^{-1} (S_{C,L}) \subset {\mathsf J}^1 \mathcal{P} F$, and assume, as a minimal regularity requirement, that $\Phi^{-1} (C )$ is a smooth subbundle of ${\mathsf J}^1 G \rightarrow G$. Then $L$ can be pulled-back to a Lagrangian density $\Phi^\ast (L) := L \circ \Phi : \Phi^{-1}(C ) \rightarrow \Omega^m$. It is easy to see that the dynamics $\mathcal{D}_{\Phi^\ast (L)} = \alpha^{-1}(S_{\Phi^{-1} (C), \Phi^\ast (L))}) \subset {\mathsf J}^1 \mathcal{P} G$ is the pre-image of $\mathcal{D}_L \subset {\mathsf J}^1 \mathcal{P} F$ under the relation $\xymatrix@C=10pt{ {\mathsf J}^1 \mathcal{P} G \ar@{.}[r] & {\mathsf J}^1 \mathcal{P}F}$. In particular, if $\Phi$ is a surjective submersion, then a section $\Sigma$ of $\mathcal{P} G$ is a solution of $\mathcal{D}_{\Phi^\ast (L)}$ iff it is related to a (necessarily unique) solution $\Phi(\Sigma)$ of $\mathcal{D}_L$ via the relation $\xymatrix@C=10pt{ \mathcal{P} G \ar@{.}[r] & \mathcal{P}F}$. On the other hand, every solution of $\mathcal{D}_L$ is (locally) of the form $\Phi (\Sigma)$ for some solution $\Sigma$ of $\mathcal{D}_{\Phi^\ast (L)}$. In other words, solutions of $\mathcal{D}_{\Phi^\ast (L)}$ project surjectively to solutions of $\mathcal{D}_L$ (up to global topological obstructions). In this sense, the dynamics $\mathcal{D}_{\Phi^\ast (L)}$ \emph{covers} the dynamics $\mathcal{D}_L$. Notice that if $\Phi = \zeta_{l,k}$, then it is a surjectve submersion (actually a fiber bundle) and the above considerations apply. This clarifies the relationship between
 \begin{itemize}
\item the dynamics generated in ${\mathsf J}^1 \mathcal{P}_k$ by a higher order Lagrangian density $L : {\mathsf J}^{k+1} \rightarrow \Omega^m$, and
\item the dynamics generated in ${\mathsf J}^1 \mathcal{P}_l$ by the same $L$ understood as a $(l+1)$-st order Lagrangian,
\end{itemize}
$l \geq k$.

Finally, notice that, in the case $\Phi = \zeta_{l,k}$, diagram (\ref{Luca12}) reduces to an obvious diagram of reduced triples, which we do not report. Similar considerations as above hold for the dynamics.

\section{Examples}

\begin{example}\label{ur:12}{\rm As a first example let us consider ``second order mechanics'', i.e.~the special case when $M={\mathbb R}$, $E=Q\times{\mathbb R}$, $\zeta$ is the projection onto the second factor, and $k=2$ (higher order mechanics, when $k > 2$, doesn't look significantly different and details about it are left to the reader). We have then ${\mathsf J}^k\simeq {\mathsf T}^kQ\times{\mathbb R}$, $\Omega^1\simeq {\mathbb R}\times {\mathbb R}$. If, moreover, $L$ does not depend explicitly on time, i.e.~it is just a function $L:{\mathsf T}^2Q\rightarrow {\mathbb R}$, we can simplify the triples dropping the factor ${\mathbb R}$ everywhere. In the ``unreduced approach'' we understand $L$ as a function on the submanifold ${\mathsf T}^2 Q\subset{\mathsf T}{\mathsf T} Q$. The unreduced Tulczyjew triple in this case is
\[
\begin{array}[c]{c}
 \xymatrix@C-20pt{
& {\mathsf T}^\ast{\mathsf T}^\ast{\mathsf T} Q\ar[dl]\ar[ddr]|!{[drr];[dl]}\hole & & & {\mathsf T}{\mathsf T}^\ast{\mathsf T} Q\ar[lll]_{\beta_{{\mathsf T} Q}}\ar[rrr]^{\alpha_{{\mathsf T} Q}} \ar[dl]\ar[ddr]|!{[drr];[dl]}\hole
& & & {\mathsf T}^\ast{\mathsf T}{\mathsf T}  Q \ar[dl]\ar[ddr]& \\
{\mathsf T}^\ast{\mathsf T} Q\ar[ddr] & & & {\mathsf T}^\ast{\mathsf T} Q\ar@{=}[lll]\ar@{=}[rrr]\ar[ddr] & & & {\mathsf T}^\ast{\mathsf T} Q \ar[ddr]& & \\
& & {\mathsf T}{\mathsf T} Q\ar[dl] & & & {\mathsf T}{\mathsf T} Q\ar@{=}[lll]|!{[ull];[dl]}\hole  \ar@{=}[rrr]|!{[ur];[drr]}\hole \ar[dl] & & & {\mathsf T}{\mathsf T} Q\ar[dl]  \\
& {\mathsf T} Q & & &{\mathsf T} Q\ar@{=}[lll]\ar@{=}[rrr] & & & {\mathsf T} Q &
}
\end{array},
\]
where $\alpha_{{\mathsf T} Q}$ is the ``Tulczyjew morphism'' for ${\mathsf T} Q$ and, similarly, $\beta_{{\mathsf T} Q}$ is the morphism determined by the canonical symplectic form $\omega_{{\mathsf T} Q}$ on ${\mathsf T}^\ast{\mathsf T} Q$. Starting form local coordinates $(q^i)$ in $Q$ and $(q^i, v^j)$ in ${\mathsf T} Q$ we get natural coordinates
$$\begin{array}{ll}\vspace{5pt}
(q^i, v^j, \dot q^i, \dot v^j) & \text{in }{\mathsf T}{\mathsf T} Q ,\\ \vspace{5pt}
(q^i, v^j, p_i, r_j) & \text{in }{\mathsf T}^\ast{\mathsf T} Q,\\ \vspace{5pt}
(q^i, v^j, \dot q^i, \dot v^j, \pi_i, \rho_j, \dot \pi_i, \dot\rho_j) & \text{in }{\mathsf T}^\ast{\mathsf T}{\mathsf T} Q, \\ \vspace{5pt}
(q^i, v^j, p_i, r_j,\dot q^i, \dot v^j, \dot p_i, \dot r_j) & \text{in }{\mathsf T}{\mathsf T}^\ast{\mathsf T} Q, \\ \vspace{5pt}
(q^i, v^j, p_i, r_j,\varphi_i, \bar\varphi_i, \psi^i, \bar\psi^j) & \text{in }{\mathsf T}^\ast{\mathsf T}^\ast{\mathsf T} Q.
\end{array}$$
It is easy to see that in coordinates
$$\begin{array}{l} \vspace{5pt}
\alpha_{{\mathsf T} Q}(q^i, v^j, p_i, r_j,\dot q^i, \dot v^j, \dot p_i, \dot r_j)=(q^i, v^j, \dot q^i, \dot v^j, \dot p_i, \dot r_j, p_i, r_j), \\
\beta_{{\mathsf T} Q}(q^i, v^j, p_i, r_j,\dot q^i, \dot v^j, \dot p_i, \dot r_j)= (q^i, v^j, p_i, r_j, -\dot p_i, -\dot r_j,\dot q^i, \dot v^j).
\end{array}$$
The submanifold ${\mathsf T}^2 Q$ in ${\mathsf T}{\mathsf T} Q$ is given by the condition $v^i=\dot q^i$. A second order Lagrangian is thus a function $L=L(q^i, v^i, \dot v^i)$. A Lagrangian function defined on ${\mathsf T}^2 Q$ generates the following (Lagrangian) submanifold in ${\mathsf T}^\ast{\mathsf T}{\mathsf T} M$
\begin{align*}
& S_{{\mathsf T}^2 Q, L} \\
&  = \left\{\,(q^i, v^j, \dot q^i, \dot v^j, \pi_i, \rho_j, \dot \pi_i, \dot\rho_j):\
v^i=\dot q^i,\  \pi_j=\frac{\partial L}{\partial q^j},\ \dot \pi_k+\rho_k=\frac{\partial L}{\partial v^k},\  \dot\rho_l=\frac{\partial L}{\partial \dot v^l}\,\right\}.
\end{align*}
The dynamics $\mathcal{D}_2=\alpha_{{\mathsf T} Q}^{-1}(S_{{\mathsf T}^2 Q, L})$ is then
\begin{align*}
& \mathcal{D}_2 \\
& =\left\{\, (q^i, v^j, p_i, r_j,\dot q^i, \dot v^j, \dot p_i, \dot r_j):\
v^i=\dot q^i,\  \dot p_j=\frac{\partial L}{\partial q^j},\  p_k+\dot r_k=\frac{\partial L}{\partial v^k},\   r_l=\frac{\partial L}{\partial \dot v^l}\,\right\}.
\end{align*}
In general, $\mathcal{D}_2$ is an implicit differential equation (imposed on curves in ${\mathsf T}^\ast{\mathsf T} Q$) and the Hamiltonian generating object is a family
$$F:{\mathsf T}^\ast{\mathsf T} Q\times_{{\mathsf T} Q}{\mathsf T}^2 Q \longrightarrow {\mathbb R}, \quad (q^i, v^j, p_i, r_j, \dot v^k)\longmapsto p_iv^i+r_j\dot v^j-L(q^i, v^i, \dot v^i).$$

For instance, let $L$ be the Lagrangian governing the motion of the tip of a javelin \cite{C}. The manifold of positions is $Q={\mathbb R}^3$  and
\begin{equation}\label{ex:6}
L(q^i, v^i, \dot v^i)=\frac12\sum_{i=1}^3(v^i)^2-(\dot v^i)^2 .
\end{equation}
The dynamics is given is
$$ \mathcal{D}_2 : \quad v^i=\dot q^i,\  \dot p_j=0,\  p_k+\dot r_k=v^k,\   r_l=-\dot v^l.$$
It is easy to see that $\mathcal{D}_2$ is the image of the vector field
$$X(q^i, v^j, p_i, r_j)=v^i\frac{\partial}{\partial q^i}-r_j\frac{\partial}{\partial v^j}+(v^k-p_k)\frac{\partial}{\partial r_k}.$$
Accordingly, the Hamiltonian generating object
$$F:{\mathsf T}^\ast{\mathsf T} Q\times_{{\mathsf T} Q}{\mathsf T}^2 Q \longrightarrow {\mathbb R}, \  (q^i, v^j, p_i, r_j, \dot v^k)\longmapsto p_kv^k+r_j\dot v^j-\frac12\sum_{i=1}^3(v^i)^2-(\dot v^i)^2, $$
can be simplified. Namely, the condition for a critical point can be solved for $\dot v^i$:
$$\frac{\partial F}{\partial \dot v^i}=r_i+\dot v^i=0\quad\Longrightarrow\quad \dot v^i=-r_i, $$
and the dynamics is generated by one single Hamiltonian function which reads
$$H:{\mathsf T}^\ast{\mathsf T} Q \longrightarrow {\mathbb R}, \quad  (q^i, v^j, p_i, r_j) \longmapsto p_kv^k-\frac{1}{2}\sum_{i=1}^3(r_i)^2+(v^i)^2.$$
Interestingly enough, in this example, although the Lagrangian generating object is a function defined on a submanifold, nonetheless the dynamics is an explicit differential equation given by a Hamiltonian vector field. $\diamondsuit$
}\end{example}

\begin{example}\label{ex:2} In the reduced triple approach to second order mechanics the ``Lagrangian space'' is $\operatorname{Lag}_1 ={\mathsf T}^\ast{\mathsf T}^2 Q$. The Lagrangian space is a double bundle with vector bundle structure over ${\mathsf T}^2 Q$ and affine bundle structure over $\mathcal{Q}_1={\mathsf T} Q\times_Q{\mathsf T}^\ast Q$. The Hamiltonian bundle $\operatorname{Ham}_1={\mathsf P}({\mathsf T}^2 Q)^\dag$ is constructed from the one dimensional affine bundle $\mathcal K_1=({\mathsf T}^2Q)^\dag\rightarrow \mathcal{Q}_1$.
The reduced triple for second order mechanics is
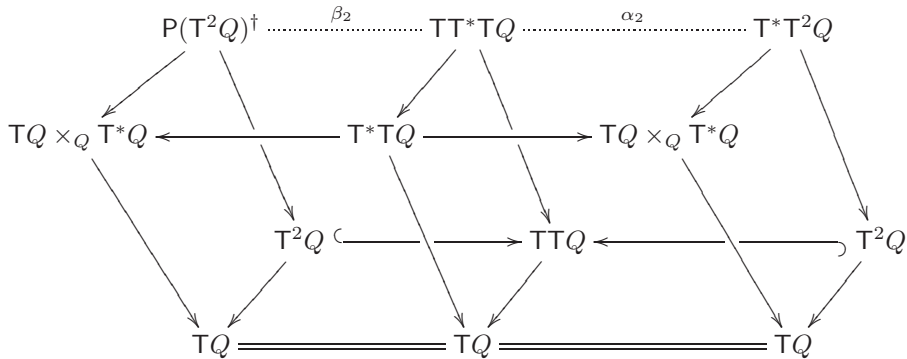
\[
\begin{array}[c]{c}
\xymatrix@C-25pt{
& {\mathsf P}({\mathsf T}^2Q)^\dag\ar[dl]\ar[ddr]|!{[drr];[dl]}\hole & & & {\mathsf T}{\mathsf T}^\ast{\mathsf T} Q\ar@{.}[lll]_{\beta_{2}}\ar@{.}[rrr]^{\alpha_{2}} \ar[dl]\ar[ddr]|!{[drr];[dl]}\hole
& & & {\mathsf T}^\ast{\mathsf T}^2 Q \ar[dl]\ar[ddr]& \\
{\mathsf T} Q\times_Q{\mathsf T}^\ast Q\ar[ddr] & & & {\mathsf T}^\ast{\mathsf T} Q\ar[lll]\ar[rrr]\ar[ddr] & & & {\mathsf T} Q\times_Q{\mathsf T}^\ast Q \ar[ddr]& & \\
& & {\mathsf T}^2 Q\ \ar[dl]\ar@{^{(}->}[rrr]|!{[ur];[drr]}\hole & & & {\mathsf T}{\mathsf T} Q\ar[dl] \ar@{-}[rrr]+<-3.5ex,0ex>|!{[ur];[drr]}\hole \ar@{-}[lll]+<3.5ex,0ex>|!{[ull];[dl]}\hole
& & &\  {\mathsf T}^2 Q\ar[dl]\ar@{^{(}->}[lll]|!{[ull];[dl]}\hole  \\
& {\mathsf T} Q & & &{\mathsf T} Q\ar@{=}[lll]\ar@{=}[rrr] & & & {\mathsf T} Q &
}
\end{array}.
\]
In local coordinates
$$\begin{array}{ll}\vspace{5pt}
(q^i,\dot q^i, \ddot q^j) & \text{in }{\mathsf T}^2 Q ,\\ \vspace{5pt}
(q^i, v^j, \dot q^i, \dot v^j) & \text{in }{\mathsf T}{\mathsf T} Q ,\\ \vspace{5pt}
(q^i, v^j, p_i, r_j) & \text{in }{\mathsf T}^\ast{\mathsf T} Q,\\ \vspace{5pt}
(q^i,\dot q^i, \ddot q^i, \pi_j, \dot \pi_j, \ddot \pi_j) & \text{in }{\mathsf T}^\ast{\mathsf T}^2 Q, \\ \vspace{5pt}
(q^i, v^j, p_i, r_j,\dot q^i, \dot v^j, \dot p_i, \dot r_j) & \text{in }{\mathsf T}{\mathsf T}^\ast{\mathsf T} Q, \\ \vspace{5pt}
(q^i, \dot q^i, r_i, \varphi_ j, \psi_j, \vartheta^j) & \text{in }{\mathsf P}({\mathsf T}^2 Q)^\dag.
\end{array}$$
we have
$$\begin{array}{l}\vspace{5pt}
{\mathsf T}^\ast{\mathsf T} Q\longrightarrow {\mathsf T} Q\times_Q{\mathsf T}^\ast Q, \quad (q^i, v^j, p_i, r_j) \longmapsto (q^i, \dot q^j=v^j, r_j), \\ \vspace{5pt}
{\mathsf T}^\ast{\mathsf T}^2 Q \longrightarrow {\mathsf T} Q\times_Q{\mathsf T}^\ast Q, \quad (q^i,\dot q^i, \ddot q^i, \pi_j, \dot \pi_j, \ddot \pi_j)\longmapsto (q^i, \dot q^j, r_j=\ddot \pi_j), \\ \vspace{5pt}
{\mathsf T}^2 Q \longrightarrow {\mathsf T}{\mathsf T} Q, \quad (q^i,\dot q^i, \ddot q^j)\longmapsto (q^i, v^j=\dot q^j, \dot q^i, \dot v^j).
\end{array}$$
In coordinates the Lagrangian relation $\alpha_1$ between $(q^i, v^j, p_i, r_j,\dot q^i, \dot v^j, \dot p_i, \dot r_j)$ and
$(q^i,\dot q^i, \ddot q^i, \pi_j, \dot \pi_j, \ddot \pi_j)$ is given by conditions
\begin{equation*}
v^i=\dot q^i, \quad \dot v^i=\ddot q^i,\quad p_j+\dot r_j=\dot\pi_j, \quad \dot p_j=\pi_j.
\end{equation*}
A point $(q^i, \dot q^i, r_i, \varphi_ j, \psi_j, \vartheta^j) \in {\mathsf P} ({\mathsf T}^2 Q)^\dag$ is in the relation $\beta_1$ with a point
\[
(q^i, v^j, p_i, r_j,\dot q^i, \dot v^j, \dot p_i, \dot r_j) \in {\mathsf T} {\mathsf T}^\ast {\mathsf T} Q
\]
 if
\begin{equation*}
\dot q^i=v^i,\quad \dot v^i=-\vartheta^i, \quad \ddot \pi_j=\dot p_j=\varphi_j, \quad p_j+\dot r_j=\psi_j.
\end{equation*}
The dynamics generated by a general Lagrangian $L:{\mathsf T}^2 Q\rightarrow {\mathbb R}$ has a family of sections of $({\mathsf T}^2 Q)^\dag\rightarrow \mathcal{Q}_1$ as Hamiltonian generating object. This generating family corresponds to a family of functions on $({\mathsf T}^2 Q)^\dag$
\begin{equation*}
F: ({\mathsf T}^2 Q)^\dag\times_{{\mathsf T} Q}{\mathsf T}^2 Q \longrightarrow {\mathbb R}, \quad (\varphi, {\mathsf T}^2\gamma)\longmapsto \varphi({\mathsf T}^2\gamma)-L({\mathsf T}^2\gamma).
\end{equation*}
In coordinates
$$F(q^i,\dot q^i, r_i,\rho,\ddot q^i)=r_i\ddot q^i+\rho-L(q^i, \dot q^i, \ddot q^i),$$
where $\rho$ is the (affine) fiber coordinate in $({\mathsf T}^2 Q)^\dag\rightarrow \mathcal{Q}_1$. In coordinates, the corresponding family of sections
$$H_F:{\mathsf T}^2 Q\times_{{\mathsf T} Q}{\mathsf T}^\ast Q\ni \rightarrow ({\mathsf T}^2 Q)^\dag$$
reads
$$\rho=L(q^i,\dot q^i,\ddot q^i)-r_j\ddot q^j.$$
When $L$ is given by (\ref{ex:6}) we get the family of sections
$$(q^i,\dot q^i, \ddot q^i, r_i)\longmapsto \left(q^i,\dot q^i, \ddot q^i, r_i, \rho=\sum_{i=1}^3\left(\frac12(\dot q^i)^2-\frac12(\ddot q^i)^2-r_i\ddot q^i\right)\right)$$
which can be reduced to one single generating section
$$ H:{\mathsf T} Q\times_Q{\mathsf T}^\ast Q \longrightarrow ({\mathsf T}^2 Q)^\dag, \  (q^i,\dot q^i, r_i)\longmapsto \left(q^i,\dot q^i, r_i, \rho=\frac12\sum_{i=1}^3\left((\dot q^i)^2+(r_j)^2\right)\right).$$
We conclude that, in the reduced triple approach, the dynamics of the tip of a javelin is generated by a Lagrangian function and a Hamiltonian section. This is an example of system with regular Lagrangian (see, e.g., \cite{Vi,Vi2}). As usual, the Legendre relation $\xymatrix@C=10pt{\lambda_2 : {\mathsf T}^2 Q\ar@{.}[r] & {\mathsf T}^\ast{\mathsf T} Q}$ is  not a map. Nonetheless, $$\ell_2:{\mathsf T}^2 Q\rightarrow {\mathsf T} Q\times_Q{\mathsf T}^\ast Q$$ (see diagram (\ref{r:11})) is a diffeomorphism which in coordinates reads
$$(q^i,\dot q^i, \ddot q^i)\longmapsto (q^i, \dot q^i, r_i=-\ddot q^i).$$
$\diamondsuit$
\end{example}

\begin{example} Most of the physical systems that can be described within Lagrangian or Hamiltonian formalisms are of order one, i.e.~their Lagrangians depend on first derivatives of configurations only (a noteworthy exception is the Einstein-Hilbert Lagrangian in General Relativity, but, even so, Einstein field equations can be equivalently derived from first order variational principles). Dependences on higher order jets appear usually as a result of idealizations in mathematical modelling. This is precisely the case of \emph{plate theory}, i.e.~the theory of thin layers of elastic material. The theory is obtained from continuum mechanics (in Lagrangian description) by assuming that one of the dimensions of the elastic body is infinitesimally small.

The main ingredients are two Riemannian manifolds $(M,\gamma)$ and $(N, g)$ where $\dim M=2$ and $\dim N=3$. The manifold $M$ is the \emph{material space}, and $N$ is the \emph{physical space}. A position of the plate in the physical space is given by a smooth immersion $\sigma: M\rightarrow N$, which can be understood as a section of the bundle $pr_M: E=M\times N\rightarrow M$. The Lagrangian is the internal energy of the elastic plate. In the present case of an infinitesimally thin plate, it depends on second order jets of the immersion $\sigma: M\rightarrow N$ via the extrinsic curvature of the surface $\sigma(M)$ in $N$ with respect to the metric $g$. This means that the space of infinitesimal configurations is ${\mathsf J}^2={\mathsf J}^2(M,N)$ (or, more precisely, the open subset of ${\mathsf J}^2$ consisting of jets of immersions). Details about how to implement the \emph{infinitesimal thickness limit} can be found in \cite{Kij}.

Let us now review all the spaces appearing in the reduced triple in the present case of plate theory. In the following, we shall assume, as usual, that $M$ is orientable. We also fix the volume form $\eta$ defined by the metric $\gamma$ and use it to identify $\Omega^2(M)$ with $M \times {\mathbb R}$. The first jet prolongation of a map $\sigma : M \rightarrow N$ identifies with the differential $df : TM \rightarrow TN$. Accordingly, the fibre of ${\mathsf J}^1={\mathsf J}^1 E\rightarrow E$ over a point $(x,y)\in M\times N$ identifies with ${\mathsf T}^\ast_x M\otimes {\mathsf T}_yN$. Abusing the notation we shall write ${\mathsf J}^1 \simeq {\mathsf T}^\ast M\otimes {\mathsf T} N$. The phase space is
$\mathcal{P}_1={\mathsf V} ^\ast {\mathsf J}^1 \otimes_{{\mathsf J}^1 }\Omega^2(M)\simeq {\mathsf V} ^\ast({\mathsf T}^\ast M\otimes {\mathsf T} N)$. In the reduced triple formulation ``the highest momenta'' are elements of $\mathcal{Q}_1=\vee^2{\mathsf T} M\otimes_{{\mathsf J}^1}{\mathsf V} ^\ast E\otimes_{{\mathsf J}^1}\Omega^2(M)$ which in this case can be expressed as
$$\mathcal{Q}_1={\mathsf J}^1\times_{(N\times M)}(\vee^2{\mathsf T} M\otimes{\mathsf T}^\ast N).$$
A point in the second factor is naturally interpreted as {\it bending moment}. Thus, it follows from the Lagrangian side of the reduced triple that the bending moment, although usually defined in coordinates, has indeed a geometric meaning.

Let us write phase equations in coordinates. Using coordinates $(x^i)$ in $M$ and $(u^\alpha)$ in $N$ we get
coordinates $(x^i, u^\alpha, u^\beta_j)$ in ${\mathsf J}^1$ and $(x^i, u^\beta_J)$ in ${\mathsf J}^2$ with $|J|\leq 2$. In the phase space we have coordinates
$(x^i, u^\alpha_j, p^k_\beta, p^{lm}_\mu)$ such that an element of $\mathcal{P}_1$ is $p=p^k_\beta{\mathrm d} u^\beta\otimes \eta_k+p^{lm}_\mu{\mathrm d} u^\mu_l\otimes \eta_m$, where $\eta_k$ is the contraction of $\eta$ with $\frac{\partial}{\partial x^k}$. The phase equations are
\begin{align*}
&\partial_i u^\alpha  = u^\alpha_i, \ \partial_i u^\alpha_j = u^\alpha_{i,j}, \\
& \partial_j p^{1j}_\alpha+p^1_\alpha=\frac{\partial L}{\partial u^\alpha_{(1,0)}}, \
\partial_j p^{2j}_\alpha+p^2_\alpha=\frac{\partial L}{\partial u^\alpha_{(0,1)}}, \\
& p^{11}_\alpha=\frac{\partial L}{\partial u^\alpha_{(2,0)}}, \  p^{22}_\alpha=\frac{\partial L}{\partial u^\alpha_{(0,2)}},\
p^{12}_\alpha+p^{21}_\alpha=\frac{\partial L}{\partial u^\alpha_{(1,1)}} \end{align*}

and can be generated, as usual, by a family of functions on $\mathcal{K}_1=({\mathsf J}^2)^\dag$ parameterized by elements of ${\mathsf J}^2$:
$$F: \mathcal{K}_1\times_{{\mathsf J}^1}{\mathsf J}^2\rightarrow {\mathbb R}, \qquad F(\varphi, {\mathsf j}^2\sigma(x))=\varphi({\mathsf j}^2\sigma(x))-L({\mathsf j}^2\sigma(x)).$$
\end{example}

\section{Tulczyjew triples on infinite jets}

So far we defined Tulczyjew triples involving an arbitrary but finite number of derivatives of the fields. There is a formal, geometric way to account for all (arbitrarily high) derivatives of the fields at the same time, which consists in using infinite jet spaces. Let us first recall the definition of infinite jets. There is a tower of fiber bundles
\begin{equation}
M \overset{\zeta}{\longleftarrow} E \overset{\zeta_{1,0}}{\longleftarrow}  {\mathsf J}^1 \longleftarrow \cdots \overset{\zeta_{k,k-1}}{\longleftarrow}  {\mathsf J}^k \overset{\zeta_{k+1,k}}{\longleftarrow} {\mathsf J}^{k+1}  \longleftarrow \cdots . \label{Luca7}
\end{equation}
The set theoretic inverse limit of sequence (\ref{Luca7}) is denoted by ${\mathsf J}^\infty E$ (or, shortly, ${\mathsf J}^\infty$) and it is called the \emph{space of infinite jets of sections of $E$}. Equivalently, ${\mathsf J}^\infty$ is the set of equivalence classes of tangency of sections of $E$ up to order $\infty$ at arbitrary points of $M$. Namely, recall that two sections of $E$ are \emph{tangent up to order $\infty$} at $x \in M$ if their local descriptions in bundle coordinates have the same partial derivatives at $x$ up to arbitrarily high order.  Tangency up to order $\infty$ is a well defined equivalence relation. The equivalence class of section $\sigma$ is denoted by ${\mathsf j}^\infty \sigma (x)$ and it is called \emph{the $\infty$-th jet of $\sigma$ at $x$}. It contains a full, intrinsic information about all derivatives of $\sigma$ at $x$. We have
\[
{\mathsf J}^\infty := \{ {\mathsf j}^\infty \sigma (x) : \sigma \text{ a local section of $E$ and $x \in M$}\}
\]
There are obvious projections $\zeta_\infty : {\mathsf J}^\infty \rightarrow M$, ${\mathsf j}^\infty \sigma (x) \mapsto x$, and, $\zeta_{\infty,l} : {\mathsf J}^\infty \rightarrow {\mathsf J}^l $, ${\mathsf j}^\infty \sigma (x) \mapsto {\mathsf j}^l \sigma (x)$. Clearly, $\zeta_\infty = \zeta_l \circ \zeta_{\infty,l} $, and $\zeta_{\infty,l} = \zeta_{p,l} \circ \zeta_{\infty,p}$, $l \leq p $.
The $\infty$-th jet space is a countable dimensional manifold which can be coordinatized as follows. Let $\mathcal{U}$ be a coordinate domain in $E$ and $(x^i , u^\alpha)$ bundle coordinates in it. There are \emph{jet coordinates} $(x^i,u^\alpha_I)$, $|I|{}< \infty$, on $\zeta_{\infty,0}^{-1}(\mathcal{U})$. Namely, pick ${\mathsf j}^\infty \sigma (x) \in \zeta_{\infty,0}^{-1}(\mathcal{U})$, and let $\sigma$ be locally given by (\ref{Luca1}). Then put
\[
x^i ( {\mathsf j}^\infty \sigma (x)) := x^i (x), \quad \text{and} \quad u^\alpha_I ({\mathsf j}^\infty \sigma (x)) := \frac{\partial^{|I|} f_1^\alpha}{\partial x^I} (x^i (x)).
\]
Notice that sequence (\ref{Luca7}) gives rise to a pull-back sequence of algebra monomorphisms
\begin{equation}
C^\infty(M) \overset{\zeta^\ast}{\longrightarrow} C^\infty(E) \longrightarrow \cdots \longrightarrow  C^\infty({\mathsf J}^k) \overset{\zeta_{k+1,k}^\ast} \longrightarrow C^\infty({\mathsf J}^{k+1})  \longrightarrow \cdots . \label{Luca8}
\end{equation}
 By definition, the \emph{algebra $C^\infty ({\mathsf J}^\infty)$ of smooth functions over ${\mathsf J}^\infty$} is the direct limit of sequence (\ref{Luca8}). In other words a smooth function on ${\mathsf J}^\infty$ is just a function on some finite jet space. Despite ${\mathsf J}^\infty$ is not a finite dimensional smooth manifold, there is a nice differential calculus on it, and one can do differential geometry on infinite jets, to a large extent. For instance, one can define vector fields and differential forms on ${\mathsf J}^\infty$ in purely algebraic terms starting from the algebraic properties of the algebra $C^\infty ({\mathsf J}^\infty)$. The interested reader can find details in \cite{B}. Since, for what concerns our purposes, the differential geometry of infinite jets does not differ much from differential geometry of finite dimensional manifolds, we will treat ${\mathsf J}^\infty$ as a standard manifold in the following, without insisting on unessential, technical details.

The maps $\zeta_\infty$ and $\zeta_{\infty, l}$ are fiber bundles (with infinite dimensional fibers), and a section $\sigma$ of $E$ can be prolonged to a section ${\mathsf j}^\infty \sigma : M \rightarrow {\mathsf J}^\infty$, $x \mapsto {\mathsf j}^\infty \sigma (x)$, called the \emph{$\infty$-th jet prolongation of $\sigma$}. If $\sigma$ is locally given by (\ref{Luca1}), then ${\mathsf j}^\infty \sigma$ is locally given by:
\[
{\mathsf j}^\infty \sigma : u^\alpha_I = \frac{\partial^{|I|} f^\alpha}{\partial x^I} (x^i),\quad |I|{}< \infty,
\]
and contains a full, intrinsic information about all derivatives of $\sigma$. Sections of ${\mathsf J}^\infty$ of the form ${\mathsf j}^\infty \sigma$ are called \emph{holonomic sections}.

The main geometric structure on ${\mathsf J}^\infty$ consists in a canonical section $\mathcal{C}: {\mathsf J}^{\infty} \rightarrow {\mathsf J}^1 {\mathsf J}^{\infty}$ of the bundle ${\mathsf J}^1 {\mathsf J}^\infty \rightarrow {\mathsf J}^\infty$. By definition, $\mathcal{C}({\mathsf j}^{\infty} \sigma (x)) = {\mathsf j}^1 ({\mathsf j}^{\infty} \sigma ) (x)$. Since sections of the bundle ${\mathsf J}^1 {\mathsf J}^\infty \rightarrow {\mathsf J}^\infty$ are Ehresmann connections in ${\mathsf J}^\infty$ (see Section \ref{sec:2}), then $\mathcal{C}$ can be interpreted as a canonical connection in ${\mathsf J}^\infty$, sometimes called the \emph{Cartan connection}. The Cartan connection is able to \emph{detect holonomic sections of ${\mathsf J}^\infty$} in the following sense: \emph{a section $\Sigma$ of ${\mathsf J}^\infty$ is holonomic iff it is an \emph{integral section} of $\mathcal{C}$, i.e.~$j^1 \Sigma$ takes values in the image of $\mathcal{C}$}. Obviously, the Cartan connection is the $\infty$-th order analogue of the embeddings ${\mathsf J}^{k+1} \subset {\mathsf J}^1 {\mathsf J}^k$. Howevere, the latter are not connections. Because of this special feature of infinite jets, the geometry of ${\mathsf J}^\infty$ is, in many respects, much simpler than the geometry of finite jets. For instance, as we will see in a moment, the infinite order (both unreduced and reduced) Tulczyjew triples have very simple descriptions.

Let us start with the unreduced triple. The main point here is that infinitesimal configurations are first jets of sections of ${\mathsf J}^\infty$. Accordingly, the unreduced infinite order Tulczyjew triple is
\begin{equation}\label{Luca9}
\begin{array}[c]{c}
 \xymatrix@!C=1pc{
 & \operatorname{Ham}{\mathsf J}^\infty\ar[dl]\ar[ddr]|!{[dl];[drr]}\hole & & & {\mathsf J}^1\mathcal{P}{\mathsf J}^\infty \ar[lll]_-{\beta}\ar[rrr]^-{\alpha}\ar[dl]\ar[ddr]|!{[dl];[drr]}\hole & & &
 \operatorname{Lag}{\mathsf J}^\infty \ar[dl]\ar[ddr] & \\
\mathcal{P}{\mathsf J}^\infty\ar[ddr] & & & \mathcal{P}{\mathsf J}^\infty\ar@{=}[rrr]\ar@{=}[lll]\ar[ddr] & & & \mathcal{P}{\mathsf J}^\infty \ar[ddr] & & \\
    & & {\mathsf J}^1{\mathsf J}^\infty\ar@{=}[rrr]|!{[ur];[drr]}\hole\ar[dl] & & & {\mathsf J}^1{\mathsf J}^\infty\ar[dl] & & & {\mathsf J}^1{\mathsf J}^\infty \ar[dl]\ar@{=}[lll]|!{[ull];[dl]}\hole \\
 & {\mathsf J}^\infty &  & & {\mathsf J}^\infty\ar@{=}[rrr] \ar@{=}[lll] & & & {\mathsf J}^\infty \ar@/_/[ur]_-{\mathcal{C}}&
}
\end{array},
\end{equation}
where $\alpha$ and $\beta$ are defined in the usual way. The presence of the Cartan connection simplifies a lot the structure of the vertexes in (\ref{Luca9}). Indeed, the projection ${\mathsf J}^1 {\mathsf J}^\infty \rightarrow {\mathsf J}^\infty$ is an affine bundle as usual but, in addition, it possesses a distinguished section $\mathcal{C}$. Accordingly, ${\mathsf J}^1 {\mathsf J}^\infty$ identifies canonically with its model vector bundle ${\mathsf V}  {\mathsf J}^\infty \otimes_M {\mathsf T}^\ast M$. As an immediate consequence, the projection ${\mathsf J}^\dag {\mathsf J}^\infty \rightarrow \mathcal{P} {\mathsf J}^\infty$ possesses a canonical section as well, hence ${\mathsf J}^\dag {\mathsf J}^\infty$ identifies with its model vector bundle $ \mathcal{P} {\mathsf J}^\infty \times_M \Omega^m$. Finally, for similar reasons, $\operatorname{Ham} {\mathsf J}^\infty = {\mathsf P}^\dag {\mathsf J}^\dag {\mathsf J}^\infty$ possesses a canonical section and identifies with its model vector bundle ${\mathsf V} ^\ast \mathcal{P} {\mathsf J}^\infty \otimes_M \Omega^m$.

Diagram (\ref{Luca9}) plays the same role as diagram (\ref{ur:11}) for field theories depending on a not better specified number of space-time derivatives of the fields. The latter claim can be proved along very similar lines as those of Section \ref{sec:4} and we will not insist much on this. Instead, we will briefly discuss the dynamics generated by a Lagrangian density of non-specified order. Namely, if keeping track of the order of a Lagrangian density $L$ is not needed, then one can understand $L$ as an $m$-form on $M$ with values in functions on $C^\infty ({\mathsf J}^\infty)$ (recall that any smooth function on ${\mathsf J}^\infty$ is just a smooth function on some ${\mathsf J}^k$ with not better specified $k$), i.e.~a section of the bundle ${\mathsf J}^\infty \times_M \Omega^m \rightarrow {\mathsf J}^\infty $. To see a Lagrangian density $L$ as a generating object, one has to understand it as a section assigned along the submanifold ${\mathsf J}^\infty \simeq \operatorname{im}\mathcal{C} \subset {\mathsf J}^1 {\mathsf J}^\infty$. As such, it can generate a dynamics $\mathcal{D}$ in the usual way $ \mathcal{D}: = \alpha^{-1} (S_{{\mathsf J}^\infty , L}) \subset {\mathsf J}^1 \mathcal{P} {\mathsf J}^\infty$. It is easy to see that $\mathcal{D}$ does actually coincide with the Euler-Lagrange-Hamilton equations determined by $L$ \cite{Vi,Vi2}, which, in their turn, are naturally interpreted as phase equations of the theory described by $L$. One can treat in a similar way theories governed by more general generating objects.

Notice that when interpreting points in ${\mathsf J}^1 {\mathsf J}^\infty$ as infinitesimal configurations we are actually adding unphysical degrees of freedom to the theory. To get rid of them one can write down a \emph{reduced triple} for field theories of non-specified order, similarly as in Section \ref{sec:5}. However, in this case, the situation is slightly different. Indeed ${\mathsf J}^\infty$ plays both the role of ``manifold of infinitesimal configurations''and ``manifold of positions\textquotedblright . Accordingly, the projection from the former to the latter is just the identity map $\mathrm{id}: {\mathsf J}^\infty \rightarrow {\mathsf J}^\infty$, understood as a $0$-dimensional affine bundle, and the reduced triple is the rather simple diagram
\begin{equation}\label{Luca10}
\begin{array}[c]{c}
 \xymatrix@!C=1pc{
 & \operatorname{Ham}_\infty \ar[dl]\ar[ddr]|!{[dl];[drr]}\hole & & & {\mathsf J}^1\mathcal{P}{\mathsf J}^\infty \ar@{.}[lll]\ar@{.}[rrr]\ar[dl]\ar[ddr]|!{[dl];[drr]}\hole & & &
 \operatorname{Lag}_\infty \ar[dl]\ar[ddr] & \\
{\mathsf J}^\infty\ar@{=}[ddr] & & & \mathcal{P}{\mathsf J}^\infty\ar[rrr]\ar[lll]\ar[ddr] & & & {\mathsf J}^\infty \ar@{=}[ddr] & & \\
    & & {\mathsf J}^\infty\ar[rrr]|!{[ur];[drr]}\hole\ar@{=}[dl] & & & {\mathsf J}^1{\mathsf J}^\infty\ar[dl] & & & {\mathsf J}^\infty \ar@{=}[dl]\ar[lll]|!{[ull];[dl]}\hole \\
 & {\mathsf J}^\infty &  & & {\mathsf J}^\infty\ar@{=}[rrr] \ar@{=}[lll] & & & {\mathsf J}^\infty&
}
\end{array},
\end{equation}
where $\operatorname{Lag}_\infty = \operatorname{Ham}_\infty := {\mathsf V} ^\ast {\mathsf J}^\infty \otimes_M \Omega^m$. The dynamics is generated using (\ref{Luca10}) in the standard way. In particular, Lagrangian generating objects and Hamiltonian generating objects coincide if one forgets about the order of the theory.

\section{Conclusions}

Most of the physical systems that can be described within Lagrangian or Hamiltonian formalisms are of order one, i.e.~their Lagrangians depend on first derivatives of configurations only. However, idealization processes inherent to mathematical modelling, e.g.~taking infinitesimally thin layers of elastic materials, can lead to Lagrangians depending on derivatives of higher order. The Tulczyjew approach to the Lagrangian and Hamiltonian description of physical systems can be extended to such cases. Starting from the first principles of variational calculus we were able to generalize first-order Tulczyjew triple in mechanics and field theory to the higher order Tulczyjew triples of Sections \ref{sec:4} and \ref{sec:5}, providing geometric description of higher derivative field theories. Future research in this area should concentrate on applications of this general theory to particular examples.

\end{document}